\documentclass{aa}

\usepackage{txfonts}
\usepackage{graphicx}

\newcommand{\pd}[2]{\frac{\partial #1}{\partial #2}}

\begin{document}

\title{Two-component jet simulations}
\subtitle{I. Topological stability of analytical MHD outflow solutions}

\titlerunning{I. Topological stability of the self-similar solutions}

\author{T. Matsakos  \inst{1}  \and
        K. Tsinganos \inst{2}  \and
        N. Vlahakis  \inst{2}  \and
        S. Massaglia \inst{1}  \and
        A. Mignone   \inst{1,3}\and
        E. Trussoni  \inst{3}}

\authorrunning{Matsakos et al.}

\offprints{Titos Matsakos,\\
           \email{matsakos@ph.unito.it}}

\institute{Dipartimento di Fisica Generale, Universit\`a degli Studi
           di Torino, via Pietro Giuria 1, 10125 Torino, Italy    \and
           IASA and Section of Astrophysics, Astronomy and Mechanics,
           Department of Physics, University of Athens,           \\
           Panepistimiopolis, 15784 Zografos, Athens, Greece      \and
           INAF/Osservatorio Astronomico di Torino, via Osservatorio
           20, 10025 Pino Torinese, Italy}

\date{Received ?? / Accepted ??}

\abstract{Observations of collimated outflows in young stellar objects
          indicate that several features of the jets can be understood
          by adopting the picture of a two-component outflow, wherein a
          central stellar component around the jet axis is surrounded
          by an extended disk-wind.
          The precise contribution of each component may depend on the
          intrinsic physical properties of the YSO-disk system as well
          as its evolutionary stage.
          }{
          In this context, the present article starts a systematic
          investigation of two-component jet models via time-dependent
          simulations of two prototypical and complementary analytical
          solutions, each closely related to the properties of
          stellar-outflows and disk-winds.
          These models describe a meridionally and a radially
          self-similar exact solution of the steady-state, ideal
          hydromagnetic equations, respectively.
          }{
          By using the PLUTO code to carry out the simulations, the
          study focuses on the topological stability of each of the
          two analytical solutions, which are successfully extended
          to all space by removing their singularities.
          In addition, their behavior and robustness over several
          physical and numerical modifications is
          extensively examined.
          Therefore, this work serves as the starting point for the
          analysis of the two-component jet simulations.
          }{
          It is found that radially self-similar solutions (disk-winds)
          always reach a final steady-state while maintaining all their
          well-defined properties.
          The different ways to replace the singular part of the
          solution around the symmetry axis, being a first
          approximation towards a two-component outflow, lead to the
          appearance of a shock at the super-fast domain corresponding
          to the fast magnetosonic separatrix surface.
          These conclusions hold true independently of the numerical
          modifications and/or evolutionary constraints that the
          models have been undergone, such as starting with a
          sub-modified-fast initial solution or different types of
          heating/cooling assumptions.
          Furthermore, the final outcome of the simulations remains
          close enough to the initial analytical configurations showing
          thus, their topological stability.
          Conversely, the asymptotic configuration and the stability of
          meridionally self-similar models (stellar-winds) is related
          to the heating processes at the base of the wind.
          If the heating is modified by assuming a polytropic relation
          between density and pressure, a turbulent evolution is found.
          On the other hand, adiabatic conditions lead to the
          replacement of the outflow by an almost static atmosphere.}{}


\maketitle

\section{Introduction}
\label{sec:intro}

Observations made over the last two decades have shown that one
class of the widespread astrophysical phenomenon of collimated
plasma outflows (jets) is being launched from the vicinity of
most young stellar objects (YSOs) (Burrows et al. \cite{Bur96}).
These supersonic mass outflows are found to be correlated with
accretion (Cabrit et al. \cite{Cab90}; Hartigan et al.
\cite{Har95}), to have narrow opening angles (Ray et al.
\cite{Ray96}) and to propagate for several orders of magnitude of
spatial distances ranging from the AU to the pc scales (Dougados
et al. \cite{Dou00}; Hartigan et al. \cite{Har04}). A central
role in the launching, acceleration and collimation of these jets
is widely believed to be played by magnetohydrodynamic (MHD)
effects, which can also successfully remove the excessive angular
momentum, allowing in this way the YSO to accrete and enter the
main sequence. Nevertheless, although recent high angular
resolution observations put several constraints on the different
driving mechanisms proposed, it is not yet clear which is the
dominant plasma launching mechanism in YSO jets.

The system of a protostellar object basically contains two
dynamical constituents, a central protostar and its surrounding
accretion disk. Consequently, in Bogovalov \& Tsinganos
(\cite{Bog01}) it is argued that jets observed from T Tauri stars
most likely consist of two main steady components:
(i) an inner pressure driven wind, which is non-collimated if the
star is an inefficient magnetic rotator and
(ii) an outer magneto-centrifugally driven disk-wind which
provides most of the high mass loss rate observed. The relatively
faster rotating magnetized disk produces the self-collimated wind
which then forces all enclosed outflow from the central source to
be collimated as well. This conclusion is confirmed by
self-consistent simulations of the MHD equations. More recently,
in Ferreira et al. (\cite{Fer06}) it is argued that for the YSO
jets observed in association with T Tauri stars, in addition to
the pressure driven stellar outflow and the magneto-centrifugally
launched extended ``warm'' disk-wind, a third component may be
driven by magnetic processes at the magnetosphere/disk interaction,
i.e., a sporadically ejected X-type wind. In addition, in Ferreira
et al. (\cite{Fer00}) a non steady ``two-flow'' scenario was also
suggested, regarding a reconnection X- and a disk-wind.
Nevertheless, the existence of such sporadic components is not
supported by observational data as being one of the major
contributors to the steady characteristics of jets, but rather
could explain the observed variability in jet emission. On the
other hand and within the same framework, recent observations
(Edwards et al. \cite{Edw06}; Kwan et al. \cite{Kwa07}) show that
both, disk and stellar winds are mainly present in T Tauri stars,
with the dominant component being determined by the intrinsic
physical properties of the particular YSO.

From the analytical studies point of view, the complexity of the
launching and collimation mechanisms of jets have forced
researchers over the past several years to treat these two
components separately. The only available analytical MHD models
for jets are those characterized by the symmetries of radial and
meridional self-similarity (Vlahakis \& Tsinganos \cite{Vla98}).
In the former case, the solution is invariant as we look at a
constant polar angle and in the latter, as we look at a constant
spherical radius. The computational consequence of the
respective symmetry is that by employing the separable spherical
coordinates ($r$, $\theta$), the set of coupled MHD equations
reduces to a set of ordinary differential equations in $\theta$,
or, in $r$, respectively. The last remaining difficulty is to
select solutions which are causally disconnected from the source
of the outflow, i.e., those crossing the fast magnetosonic
separatrix. In this way, one may construct either radially
self-similar solutions closely related to the properties of
magneto-centrifugally driven disk-winds (Blandford \& Payne
\cite{Bla82}; Contopoulos \& Lovelace \cite{Con94}; Ferreira
\cite{Fer97}; Vlahakis et al. \cite{Vla00}, hereafter VTST00),
or  meridionally self-similar ones to address pressure driven
stellar outflows (Sauty \& Tsinganos \cite{Sau94}; Trussoni et
al. \cite{Tru97}; Sauty et al. \cite{Sau02}, hereafter STT02).

Since each self-similar symmetry corresponds to a particular
component, we adopt the following initials: ADO (Analytical Disk
Outflow) and ASO (Analytical Stellar Outflow) to refer to
radially and meridionally self-similar solutions, respectively.

Apart from the geometry, an intrinsic distinction between these
two classes concerns the treatment of the energy equation.
Nevertheless, the symmetry difference makes them complementary to
each other, since the ADO solution becomes singular at the axis,
whereas the ASO is by definition the proper one for modeling the
area close to it. In addition, the properties of the launching
region of the disk-wind, i.e., at large polar angles, are
described more naturally by the ADO model. For a recent review
on the analytical work on MHD outflows the reader is referred to
Tsinganos (\cite{Tsi07}).

On the other hand, the increase of computational power along with
the development of sophisticated numerical codes have allowed to
study the time evolution of the MHD equations, giving us new
perspectives of the physics involved. Jet launching and collimation
have been mainly investigated with the following two methodologies:
(i) by treating the disk as a boundary (Krasnopolsky et al.
\cite{Kra99}; Ouyed et al. \cite{Ouy03}; Fendt \cite{Fen06}) and
(ii) by including the disk inside the computational box (Casse \&
Keppens \cite{Cas04}; Meliani et al. \cite{Mel06}; Zanni et al.
\cite{Zan07}). The former case allows a wider range of physical
processes and mechanisms to be studied, whereas the latter, has the
advantage of the jet evolution being consistent with that of the
disk. However, with the exception of Gracia et al. (\cite{Gra06};
hereafter GVT06), most numerical studies did not take advantage of
the availability of the well studied analytical solutions which
also allow a parametric study and therefore a better physical
understanding of the problem of jet launching and collimation.

The present work is the first attempt to numerically construct
and study a two-component jet, by using as starting point the
two well studied classes of analytical self-similar solutions.
Towards this goal, in this paper we first address the question
of the topological stability of each one of these two classes
separately, before we combine them in the following paper.

Concerning the ADO model, we shall use the VTST00 analytical
solution, completing and considerably extending the GVT06 analysis.
Therein, they found that the disk-wind model may attain a new
steady-state configuration close enough to the initial analytical
one, provided the assumption of some appropriate approximations
around the axis. We further present the first numerical studies of
ASO (meridionally self-similar winds), referring to the solutions
of STT02, that are essential to model the region around the axis,
just where the ADO model fails.

Once the physical properties and stability of these two classes of
solutions is clarified, our final aim will be to effectively build
up a model that consistently merges the ASO and ADO solutions.
Such simulations will be presented in a future work where we will
study the launching and propagation of a collimated stellar wind
around the system axis surrounded by a disk wind.

Finally, a word on the term topological stability used in this
paper. Classical stability theory addresses the question whether
a given equilibrium configuration evolves away from (=unstable)
or back to (=stable) the initial equilibrium when perturbed. In
the present context, topological (or structural) stability refers
to the question whether a given configuration preserves its
topological properties when subject to various perturbations.
Needless to say, that topologically unstable configurations may
well be stable from the classical point of view and vice versa.

The paper is structured as follows: In section \S\ref{sec:anmod}
the formalism of the ADO and ASO solutions is briefly presented.
In section \S\ref{numst} their implementation is explained and
the numerical models to be investigated are presented. Section
\S\ref{sec:res} reports the results obtained by carrying out the
respective simulations. Finally in section \S\ref{sec:disc} we
discuss our results in the framework of the future matching and
we report the conclusions of this work.

\section{MHD equations and the self-similar solutions}
\label{sec:anmod}

The ideal MHD equations are:
\begin{equation}
\pd{\rho}{t} + \nabla \cdot (\rho \vec V) = 0
\,,
\label{eq:dens}
\end{equation}
\begin{equation}
\pd{\vec V}{t} + (\vec V \cdot \nabla)\vec V
+ \frac{1}{\rho}\vec B \times (\nabla \times \vec B)
+ \frac{1}{\rho}\nabla P = -\nabla \Phi
\,,
\end{equation}
\begin{equation}
\pd{P}{t} + \vec V \cdot \nabla P
+ \Gamma P \nabla \cdot \vec V = \Lambda
\,,
\label{eq:en}
\end{equation}
\begin{equation}
\pd{\vec B}{t} - \nabla \times (\vec V \times \vec B) = 0
\,,
\label{eq:induct}
\end{equation}
where $\rho$, $P$, $\vec V$, $\vec B$ denote the density,
pressure, velocity and magnetic field over $\sqrt{4 \pi}$,
respectively. $\Phi=-\mathcal{GM}/R$ is the gravitational
potential of the central object ($\mathcal G$ is the
gravitational constant) with mass ${\mathcal M}$,
$\Lambda$ represents the volumetric energy gain/loss
terms ($\Lambda = [\Gamma - 1] \rho \mathcal Q$, with
$ \mathcal Q$ the energy source terms per unit mass),
and $\Gamma$ is the ratio of the specific heats.

For the sake of clarity, we adopt the following notation: the
subscripts $r$ and $\theta$ of the physical variables, will be
used to refer to the ADO and ASO solutions respectively,
whereas the cylindrical radial direction is denoted with the
symbol $\varpi$ and the spherical radial direction with $R$.
Finally, the index $p$ corresponds to the poloidal components
of the physical variables.

By assuming steady-state and axisymmetry, several conserved
quantities exist along the fieldlines (Tsinganos \cite{Tsi82}).
By introducing the magnetic flux function
$A = (\frac{1}{2 \pi})\int \vec B_p \cdot d \vec S$ to label
the iso-surfaces that enclose constant poloidal magnetic flux,
then these integrals take the following simple formulation:
\begin{equation}
\Psi_A(A) = \frac{\rho V_p}{B_p}
\,,
\label{eq:intp}
\end{equation}
\begin{equation}
\Omega(A) = \frac{1}{\varpi} \left( V_{\phi}
- \frac{\Psi_A B_{\phi}}{\rho} \right)
\,,
\end{equation}
\begin{equation}
L(A) = \varpi \left( V_{\phi} - \frac{B_{\phi}}{\Psi_A} \right)
\,,
\end{equation}
where $\Psi_A$ is the mass-to-magnetic-flux ratio, $\Omega$
the field angular velocity, and $L$ the total specific
angular momentum. The ratio $\sqrt{L/\Omega}$ defines the
Alfv\'enic lever arm at each fieldline, where the poloidal
flow speed is equal to the poloidal Alfv\'enic one. In the
adiabatic-isentropic case where $\Lambda = 0$, there exist
two more integrals, the total energy flux density to mass
flux density $E$ and the specific entropy $Q$, which are
given by:
\begin{equation}
E(A) = \frac{V^2}{2} + \frac{\Gamma}{\Gamma - 1} \frac{P}{\rho}
+ \Phi - \Omega \varpi \frac{B_{\phi}}{\Psi_A}
\,,
\label{eq:inte}
\end{equation}
\begin{equation}
Q(A) = \frac{P}{\rho^{\Gamma}}\,.
\end{equation}

\subsection{ADO - The radially self-similar model}

We employ the radially self-similar solution which is described
in VTST00 and crosses successfully all three critical surfaces.
We note that a polytropic relation between the density and the
pressure is assumed, i.e. $P = Q(A)\rho^{\gamma}$, with $\gamma$
being the effective polytropic index. Equivalently, the source
term in Eq. (\ref{eq:en}) has the special form
\begin{equation}
\Lambda=(\Gamma - \gamma)P(\nabla\cdot\vec{V})\,,
\label{eq:equiv}
\end{equation}
transforming the energy equation (\ref{eq:en}) to
\begin{equation}
\pd{P}{t} + \vec V \cdot \nabla P
+ \gamma P \nabla \cdot \vec V = 0
\,.
\label{eq:enpol}
\end{equation}
The latter can be interpreted as describing the adiabatic
evolution of a gas with ratio of specific heats $\gamma$,
whose entropy $P/\rho^\gamma$ is conserved.

The solution is provided by the values of the key functions
$M_r(\theta)$, $G_r(\theta)$ and $\psi_r(\theta)$, which are
the Alfv\'enic Mach number, the cylindrical distance in units
of the corresponding Alfv\'enic lever arm and the angle
between a particular poloidal fieldline and the cylindrical
radial direction, respectively. Then, the fieldlines can be
labeled by
\footnote{
Note that $x$ is a model parameter governing the scaling of
the magnetic field and is related to $\xi = 2(x-3/4)$
which is a local measure of the disk ejection efficiency
in the disk model of Ferreira (\cite{Fer97}).}
\begin{equation}
A_r = \frac{B_{r*}\varpi_*^2}{x}\alpha_r^{x/2}\,,
\quad \mathrm{where} \quad\alpha_r =
\frac{\varpi^2}{\varpi_*^2G_r^2}\,.
\label{eq:arss}
\end{equation}
Hence, one obtains the following expressions for the physical
variables:
\begin{equation}
\rho_r = \rho_{r*}\alpha_r^{x-3/2}\frac{1}{M_r^2}
\,,
\end{equation}
\begin{equation}
P_r = P_{r*}\alpha_r^{x-2}\frac{1}{M_r^{2\gamma}}
\,,
\end{equation}
\begin{equation}
\vec V_{p;r} = -V_{r*}\alpha_r^{-1/4}\frac{M_r^2}{G_r^2}
\frac{\sin\theta}{\cos(\psi_r + \theta)}
\left(\cos\psi_r \hat \varpi + \sin\psi_r \hat z\right)
\,,
\label{eq:vprss}
\end{equation}
\begin{equation}
V_{\phi;r} = V_{r*}\lambda\alpha_r^{-1/4}
\frac{G_r^2-M_r^2}{G_r(1-M_r^2)}
\,,
\end{equation}
\begin{equation}
\vec B_{p;r} = -B_{r*}\alpha_r^{x/2-1}\frac{1}{G_r^2}
\frac{\sin\theta}{\cos(\psi_r + \theta)}
\left(\cos\psi_r \hat \varpi + \sin\psi_r \hat z\right)
\,,
\label{eq:bprss}
\end{equation}
\begin{equation}
B_{\phi;r} = -B_{r*}\lambda\alpha_r^{x/2-1}
\frac{1-G_r^2}{G_r(1-M_r^2)}
\,.
\label{eq:bfrss}
\end{equation}
The starred quantities are related to their characteristic
values at the Alfv\'en radius $\varpi_*$ along the reference
fieldline $\alpha = 1$. Moreover, they are interconnected
with the following relations:
\begin{equation}
V_{r*} = \frac{B_{r*}}{\sqrt{\rho_{r*}}}, \quad
P_{r*} = \frac{\mu B_{r*}^2}{2}, \quad
\sqrt{\frac{\mathcal G \mathcal M}{\varpi_* V_{r*}^2}} =
\mathcal K
\,,
\label{eq:rssparam}
\end{equation}
where the constants $\lambda$ measures the strength of
rotation and $\mathcal K$ the gravitational potential.
Finally $\mu$ is proportional to the gas entropy.

\subsection{ASO - The meridionally self-similar model}

We employ a meridionally self-similar solution which
corresponds to the case of a spherically symmetric thermal
pressure (case $\kappa = 0$, second sub-table of Table 1 in
STT02, where $\kappa$ represents the deviations from such a
pressure symmetry). This solution is derived without the
assumption of a polytropic relation, with the respective
energy source term being consistently derived a posteriori.
In this case the key functions are $G_\theta(R)$,
$M_\theta(R)$, $\Pi_\theta(R)$ and $F_\theta(R)$. The former
two have the same interpretation as in the ADO model, while
the latter ones are the pressure and the expansion factor,
respectively. The magnetic flux $A$ is then given by
\begin{equation}
A_\theta = \frac{B_{\theta *}R_*^2}{2}\alpha_\theta\,,
\quad \mathrm{where} \quad\alpha_\theta =
\frac{\varpi^2}{R_*^2G_\theta^2}\,.
\label{eq:atss}
\end{equation}

Moreover, the physical variables are given from the following
expressions:
\begin{equation}
\rho_\theta = \rho_{\theta *}
\frac{1}{M_\theta^2}(1+\delta\alpha_\theta)
\,,
\end{equation}
\begin{equation}
P_\theta = P_{\theta *}\Pi_\theta
\,,
\end{equation}
\begin{equation}
V_{\varpi;\theta} = V_{\theta *}\frac{M_\theta^2}{G_\theta^2}
\frac{\sin\theta \cos\theta}{\sqrt{1+\delta\alpha_\theta}}
\left(1 - \frac{F_\theta}{2}\right)
\,,
\end{equation}
\begin{equation}
V_{z;\theta} = V_{\theta *}\frac{M_\theta^2}{G_\theta^2}
\frac{1}{\sqrt{1+\delta\alpha_\theta}}
\left(\cos^2\theta + \sin^2\theta\frac{F_\theta}{2}\right)
\,,
\end{equation}
\begin{equation}
V_{\phi;\theta} = V_{\theta *}\lambda'\alpha_\theta^{1/2}
\frac{G_\theta^2-M_\theta^2}{G_\theta(1-M_\theta^2)}
\frac{1}{\sqrt{1+\delta\alpha_\theta}}
\,,
\end{equation}
\begin{equation}
B_{\varpi;\theta} = B_{\theta *}
\frac{\sin\theta \cos\theta}{G_\theta^2}
\left(1 - \frac{F_\theta}{2}\right)
\,,
\end{equation}
\begin{equation}
B_{z;\theta} = B_{\theta *}\frac{1}{G_\theta^2}
\left(\cos^2\theta + \sin^2\theta\frac{F_\theta}{2}\right)
\,,
\end{equation}
\begin{equation}
B_{\phi;\theta} = -B_{\theta*}\lambda'\alpha_\theta^{1/2}
\frac{1-G_\theta^2}{G_\theta(1-M_\theta^2)}
\,.
\label{eq:bftss}
\end{equation}
Here, $\delta$ describes deviations from a
spherically-symmetric density whereas $\lambda'$ the
strength of the magnetic torque at the Alfv\'en radius
$R_*$. The starred quantities are the reference values
at $R_*$ and are related as follows:
\begin{equation}
V_{\theta*} = \frac{B_{\theta*}}{\sqrt{\rho_{\theta*}}},
\quad P_{\theta*} = \frac{1}{2}B_{\theta*}^2, \quad
\sqrt{\frac{2\mathcal G \mathcal M}{R_* V_{\theta*}^2}} = \nu
\,,
\label{eq:tssparam}
\end{equation}
where $\nu$ represents the strength of the gravitational
potential.

Finally, the expression for the energy source term is
\begin{equation}
\Lambda = \frac{\rho_\theta V_{R;\theta}}{1+\delta\alpha_\theta}
\frac{V_{\theta *}^2}{2R_*}\left(M_\theta^2\frac{d\Pi_\theta}{dR}
+ \Gamma\Pi_\theta\frac{dM_\theta^2}{dR}\right)
\,.
\label{eq:lam}
\end{equation}

\subsection{Physical aspects and differences of the two solutions}

In order to have a better understanding of these two classes
of self-similar solutions, we pause here to point out a few
aspects concerning the main physical mechanisms involved in
each case and their major intrinsic differences.

The ADO (radially self-similar) solution corresponds to a
magneto-centrifugally driven outflow with the ``bead on a
rotating wire'' analogy. Fig.~7 of VTST00 displays the
different terms of the conserved total energy as plotted along
a particular fieldline as a function of $z$. It is evident,
that close to the base of the outflow, the electromagnetic
energy dominates, whereas, as the flow is being accelerated,
the poloidal kinetic one becomes eventually the main component
of the total energy. The minor role that the enthalpy seems to
play is being investigated in \S\ref{sec:res}. Finally the
reader is also referred to Fig.~5 of VTST00 where the
components of the outflow speeds are plotted.

On the contrary, the ASO (meridionally self-similar) solution
adopted is a pressure driven outflow. Fig.~9 of STT02 presents
the forces acting along and across the streamlines for a
solution very similar to the one employed here. It can be
clearly seen that the pressure gradient is the dominant force
for the acceleration of the flow while the collimation
mechanisms are due to the hoop stresses. Another important
feature of the ASO model is that apart from the polar
fieldlines leaving the stellar surface and closing to
infinity, there are also those which cross the equator being
asymptotically parallel to the axisymmetry axis. Moreover, a
``dead-zone'' also exists and is defined by the region of the
fieldlines with both their footpoints rooted on the star
(Fig.~10 of Sauty \& Tsinganos \cite{Sau94}). Furthermore, we
note that meridionally self-similar models are classified by
an energetic criterion which characterizes the asymptotic
shape of the streamlines. In our case, the employed ASO
corresponds to a cylindrical, magnetically collimated jet.

A common feature concerning both classes, is the fact that the
poloidal critical surfaces do not coincide with the surfaces
where the steady-state MHD equations change character from
elliptic to hyperbolic and vice versa. This is because of the
constraint put on the propagation of the MHD waves by the
axisymmetry and the self-similarity assumption (Tsinganos et
al. \cite{Tsi96}).

As far as the two main intrinsic differences of the two models
are concerned, we note the following. First, due to the
assumptions of self-similarity, the ADO model has its MHD
critical surfaces given for $\theta_{cr} = const$, hence being
of conical shape. On the other hand, they are spherical in the
case of the ASO, as an outcome of the radial dependence of its
key functions.

The second difference concerns the energy equation. In the ADO
solution, by assuming a polytropic relationship between density
and pressure, the total energy-to-mass-flux-ratio is conserved.
However, in the ASO, the momentum equation provides enough
relations to close the system and the total
energy-to-mass-flux-ratio is not used. In both cases though, the
heating/cooling mechanisms necessary to maintain the outflow can
be calculated a posteriori (see Eq. [\ref{eq:equiv}] and
[\ref{eq:lam}]).

\section{The numerical models}
\label{numst}

We will mainly focus on four topics sorted by increasing importance:
\begin{enumerate}
\item
to complete the GVT06 work by imposing the correct number of
boundary conditions according to the number of waves propagating
downstream;
\item
to further extend the GVT06 results by including the equator inside
the computational domain and by investigating the effects of the
singularity substitution, the resolution and the choice of the
minimum vertical distance of the lower boundary of the computational
box, $z_{min}$;
\item
to carry out and present the first time-dependent simulations of
an ASO solution;
\item
to study how the energy input/output modifications influence
the features, stability and robustness of each model.
\end{enumerate}

The analytical solutions provide the key functions, already
discussed, along $\theta$ and $R$ for the ADO and ASO models,
respectively. Then, by properly interpolating in a cylindrical
or a spherical grid the physical values are initialized with
the help of Eqs. (\ref{eq:arss})-(\ref{eq:bfrss}) and
(\ref{eq:atss})-(\ref{eq:bftss}). More details on the treatment
of the axis concerning the ADO solution are given in section
\S\ref{sec:res}.

Eqs. (\ref{eq:dens})-(\ref{eq:induct}) are solved numerically
using the MHD module provided by the PLUTO code
\footnote{
Publicly available at \textit{http://plutocode.to.astro.it}}
(Mignone et al. \cite{Mig07}). PLUTO is a modular Godunov-type
code particularly oriented towards the treatment of astrophysical
flows in the presence of discontinuities. For the present case,
second order accuracy is achieved using a Runge-Kutta scheme (for
temporal integration) and piece-wise linear reconstruction (in
space). Although all the computations were carried out with the
simple (and computationally efficient) Lax-Friedrichs solver, we
point out that no significant differences were found by switching
to more complex Riemann solvers available in the code.

\begin{table*}
\caption{List of the numerical models. The first 14 lines refer
to Analytical Disk Outflows (ADO) solutions, the last 7 lines to
the Analytical Stellar Outflow (ASO) solutions.}
\label{table:1}
\centering
\begin{tabular}{l c l c c c l}
\hline
\hline
Name   & Model       & Geometry
       & Grid $[\varpi \times z]$ or $[R \times \theta]$
       & Resolution  & Total time  & Description                     \\
\hline
$1DR$  & ADO         & Cylindrical & $[0,50]\times[6,100]$
       & $128 \times 240$   & 6.0
       & GVT06 (overspecified b.c.)
         (Fig.~\ref{fig:integrals})                                  \\
$2DB$  & ADO         & Cylindrical & $[0,50]\times[6,100]$
       & $128 \times 240$   & 6.0
       & Correct number of b.c.
         (Fig.~\ref{fig:integrals}, \ref{fig:invar})                 \\
$3DB$  & ADO         & Cylindrical & $[0,50]\times[6,100]$
       & $256 \times 480$   & 6.0
       & Correct number of b.c., higher resolution,
         (Fig.~\ref{fig:bound})                                      \\
$4DH$  & ADO         & Cylindrical & $[0,50]\times[6,56]$
       & $400 \times 400$   & 2.0
       & Shock study                                                 \\
$5DH$  & ADO         & Cylindrical & $[0,50]\times[6,56]$
       & $800 \times 800$   & 2.0
       & Shock study
         (Fig.~\ref{fig:charac}, \ref{fig:currents})                 \\
$6DH$  & ADO         & Cylindrical & $[0,50]\times[6,56]$
       & $1200 \times 1200$ & 2.0
       & Shock study
         (Fig.~\ref{fig:shock})                                      \\
$7DS$  & ADO         & Cylindrical & $[0,50]\times[6,100]$
       & $128 \times 240$   & 6.0
       & Different singularity smoothening
         (Fig.~\ref{fig:m}, \ref{fig:invar})                         \\
$8DS$  & ADO         & Cylindrical & $[0,50]\times[6,100]$
       & $128 \times 240$   & 6.0
       & Sub modified fast lower boundary
         (Fig.~\ref{fig:m}, \ref{fig:sing}, \ref{fig:invar})         \\
$9DZ$  & ADO         & Cylindrical & $[0,50]\times[3,100]$
       & $128 \times 248$   & 15.0
       & Lower $z_{min}$
         (Fig.~\ref{fig:zmin})                                       \\
$10DZ$ & ADO         & Cylindrical & $[0,50]\times[12,100]$
       & $128 \times 224$   & 4.0
       & Higher $z_{min}$
         (Fig.~\ref{fig:zmin})                                       \\
$11DE$ & ADO         & Spherical   & $[10,90]\times[0,\pi/2-\epsilon]$
       & $408 \times 200$   & 2.0
       & Extension up to the equator
         (Fig.~\ref{fig:spher})                                      \\
$12DG$ & ADO         & Spherical   & $[10,90]\times[0,\pi/2-\epsilon]$
       & $408 \times 200$   & 10.0
       & Adiabatic, $\Gamma = 5/3$
         (Fig.~\ref{fig:spher})                                      \\
$13DI$ & ADO         & Spherical   & $[10,90]\times[0,\pi/2-\epsilon]$
       & $408 \times 200$   & 2.0
       & Isothermal
         (Fig.~\ref{fig:spher})                                      \\
$14DT$ & ADO         & Cylindrical &  $[0,200]\times[6,100]$
       & $512 \times 240$   & 220.0
       & Long term simulation
         (Fig.~\ref{fig:time})                                       \\
\hline
$1SR$  & ASO        & Cylindrical & $[0,50]\times[6,100]$
       & $128 \times 240$   & 4.0
       & Super-Alfv\'enic domain in cylindrical
         (Fig.~\ref{fig:tss}, \ref{fig:tssint})                      \\
$2SL$  & ASO        & Spherical   & $[1,1000]\times[0,\pi/2-\epsilon]$
       & $200 \times 200$   & 40.0
       & Log. grid, super-Alfv\'enic domain                          \\
$3SL$  & ASO        & Spherical   & $[0.7,7]\times[0,\pi/2-\epsilon]$
       & $200 \times 200$   & 4.0
       & Log. grid, sub-Alfv\'enic included
         (Fig.~\ref{fig:logar}, \ref{fig:heat})                      \\
$4SP$  & ASO        & Cylindrical & $[0,50]\times[6,100]$
       & $128 \times 240$   & 600.0
       & Use of polytropic index $\gamma = 1.05$
         (Fig.~\ref{fig:tssint})                                     \\
$5SG$  & ASO        & Cylindrical & $[0,50]\times[6,100]$
       & $128 \times 240$   & 600.0
       & Adiabatic, $\Gamma = 5/3$
         (Fig.~\ref{fig:tssint})                                     \\
$6SL$  & ASO        & Spherical   & $[0.7,7]\times[0,\pi/2-\epsilon]$
       & $200 \times 200$   & 80.0
       & Log. grid, trans-Alfv\'enic, polytropic
         (Fig.~\ref{fig:diff})                                       \\
$7SL$  & ASO        & Spherical   & $[0.7,7]\times[0,\pi/2-\epsilon]$
       & $200 \times 200$   & 80.0
       & Log. grid, trans-Alfv\'enic, adiabatic
         (Fig.~\ref{fig:diff})                                       \\
\hline
\end{tabular}
\end{table*}

Table~\ref{table:1} lists the numerical models constructed in
order to study the previously mentioned aspects. Note that the
ADO model is being investigated more extensively due to the
singularity appearing at small polar angles. We define the
reference lengths $\varpi_*$ and $R_*$, of the ADO and the ASO
models respectively, to be unity. In addition, the reference
velocities are normalized by setting $V_{r*} = 1$ and
$V_{\theta*} = \sqrt{2}\mathcal K / \nu$ in order for both
solutions to have the same gravitational potential. Time will
be expressed in units of
$t_* = 2\pi\sqrt{\varpi_*^3 / \mathcal G \mathcal M} =
2\pi\sqrt{\varpi_*^2 V_{r*}^2 / \mathcal K} = \pi$,
i.e. using the Keplerian period at distance $\varpi_*$ or $R_*$
on the equatorial plane. The final time of the simulations is
obviously chosen to be greater than the one needed for a
steady-state to be reached, if of course there exists one.
Notice that all the following figures, apart from
Figs.~\ref{fig:m} and \ref{fig:diff}, correspond to the final
state reached by the simulation at the final time indicated
in Table~\ref{table:1}. In order to prove the stability of
the steady state we have included a long term simulation to
make the argument solid. Finally, concerning the choice of
our computational box, in many models we follow the
guidelines of GVT06.

\subsection{The initial ADO model}

The model parameters of this solution were chosen as $x = 0.75$
and $\gamma = 1.05$, while the solution parameters are given to
be $\lambda = 11.70$, $\mu = 2.99$, $\mathcal K = 2.00$, in
accordance to VTST00
\footnote{The solution adopted with the model parameter
$x = 0.75$ corresponds to a zero ejection index $\xi$ according
to Ferreira (\cite{Fer97}). However, as it is evident from
Figs.~5 and 6 of VTST00 the solution with $x = 0.7575$, i.e.
$\xi = 0.0025$, is almost identical to the one with $x = 0.75$
for $z \gtrsim 0.1$. Therefore, we argue that the ADO solution
employed here should not contradict the theoretical arguments
presented in Ferreira (\cite{Fer97}).}.

By definition, radially self-similar models fail to provide
physically accepted solutions close to the axis, due to the
local diverging behavior of the physical variables. The solution
of VTST00 is terminated at $\theta < \theta_{min} = 0.025 (rad)
\simeq 1.5^\circ$, after having crossed the modified fast
critical surface, also known as the fast magnetosonic separatrix
surface (FMSS) (Tsinganos et al. \cite{Tsi96}). However, in GVT06
the rotational axis was included self-consistently inside the
computational box by properly initializing the physical variables
at these small polar angles. This was achieved by assuming that
the key functions $G(\theta)$, $M(\theta)$, and $\psi(\theta)$
are all even, e.g. $G(\theta)=G(-\theta)$, and then by
accordingly interpolating. The time evolution was performed with
the numerical code NIRVANA. We have implemented this setup (model
$1DR$) in PLUTO, although with a slightly different extrapolation
scheme. Our proper smoothening of the flow quantities near the
axis follows the same guidelines, i.e. linearly extrapolating the
key functions and then initializing the physical variables. This
is discussed in detail later on, where we apply and investigate
the effect of other extrapolation schemes as well. Finally,
notice that such a modification of the inner part of the wind,
mimics the presence of an effective stellar outflow.

Obviously, such assumptions would not retain a divergence-free
magnetic field in the central region, if Eq. (\ref{eq:bprss})
is used for its initialization. So, instead, we initialize the
poloidal component of the magnetic field by taking advantage of
the flux function (\ref{eq:arss}). Finally, in order to ensure
consistency with the ideal MHD assumption, the radial velocity
component is derived from the relation:
\begin{equation}
V_\varpi = B_\varpi\frac{V_z}{B_z}\,.
\label{eq:ideal}
\end{equation}
where $V_z$ is taken from Eq. (\ref{eq:vprss}).

\subsection{The initial ASO model}

The adopted analytical solution has the model parameter $\kappa = 0$,
and corresponds to the following values of the solution parameters:
$\delta = 0.01$, $\lambda' = 3$, $\nu = 24.1$, in accordance to STT02.

In this case, the values of the key functions are available for
$0.6R_* \le R \le 10^4R_*$. However, note that this model varies
significantly over all its radial scales and hence it is numerically
complicated to resolve all regions with adequate accuracy. For this
reason, a number of simulations consider the super-Alfv\'enic domain
only. On the other hand, exploiting the fact that PLUTO can
integrate in time over a uniform logarithmic grid, the full range of
the solution is also studied. Furthermore, the energy source term,
given by expression (\ref{eq:lam}), is being taken into account
during the numerical evolution, unless otherwise indicated. Notice
however, that the energy source term is not free to evolve in time,
since it is provided by the key functions of the analytical solution
and hence it is kept fixed throughout the simulations.

\subsection{Boundaries}
\label{sec:bound}

The number of boundary conditions imposed at a physical boundary has
to match the number of characteristic waves carrying information from
the boundary towards the inside of the computational box (Bogovalov
\cite{Bog97}). This defines the physical boundary conditions (Thompson
\cite{Tho87}, \cite{Tho90}; Mignone \cite{Mig05}) as opposed to the
information directed outside, which can be entirely determined by the
solution inside the computational box. Still, since our numerical
scheme requires the knowledge of all (eight) flow variables in the
ghost zones, additional numerical boundary conditions are prescribed
using suitable one-sided extrapolation formula. Furthermore, although
the physical quantities evolved in the code are $8$, i.e., $\rho$, $P$,
$\vec B$, $\vec V$, axisymmetry along with the
$\nabla \cdot \vec B = 0$ condition reduces the number of variables
to $7$.

Note that in GVT06, all the quantities were kept fixed at the lower $z$
boundary postponing the examination of this over-specification issue for
a future study.

In this framework, when cylindrical coordinates are adopted, we divide
the lower boundary in 4 regions, i.e.
a) $V_z > V_{fast;z}$, b) $V_{fast;z} > V_z > V_{Alfv\acute{e}n;z}$,
c) $V_{Alfv\acute{e}n;z} > V_z > V_{slow;z}$ and d) $V_z < V_{slow;z}$.
In region a) we keep all seven quantities ($V_\varpi$ is always taken
from Eq.~\ref{eq:ideal}) fixed to their analytical values (since all
7 waves are directed inward), while the number of extrapolated variables
increases by one each time we cross a critical surface, being left with
only 4 specified variables in d). In particular $P$, $B_\varpi$,
$B_\phi$ are free to evolve in region d), while only $P$, $B_\phi$ in
region c) and finally only $B_\phi$ in b). Note that, since the entropy
wave is always directed inside, at least four out of seven MHD waves
are associated with a physical boundary condition.

As far as the rest of the boundaries are concerned, we impose
axisymmetry at the axis and outflow conditions at the top $z$
boundary. Moreover, at the outer radial boundary, we apply outflow
conditions for the ADO, whereas we keep the derivative of
$B_\phi$ constant for the simulations of the ASO solution. This is
particularly important for the latter case where the jet shows a
high degree of collimation. As a result, a free condition for the
toroidal component of the magnetic field would cancel the poloidal
current along the boundary. Hence, the Lorentz force would be zero
and the uncompensated hoop stress would create artificial
collimation (as discussed in Zanni et al. \cite{Zan07}).

On the other hand, when spherical coordinates are adopted,
axisymmetry holds at the axis and free conditions are imposed at
the outer radial boundary. At the inner radial boundary and at
$\theta_{max}$, the correct number of conditions are specified, as
previously mentioned. Of course, this is achieved according to the
perpendicular velocities which now are $V_R$ and $V_\theta$,
respectively.

\section{Results}
\label{sec:res}

\subsection{The Analytical Disk Outflow (ADO) solution}

\subsubsection{Boundary conditions}

We begin by presenting the results obtained by the correct
specification of the boundary conditions, as discussed in
section \S\ref{sec:bound} (models $1DR$, $2DB$, $3DB$).

\begin{figure}
\resizebox{\hsize}{!}{
 \includegraphics{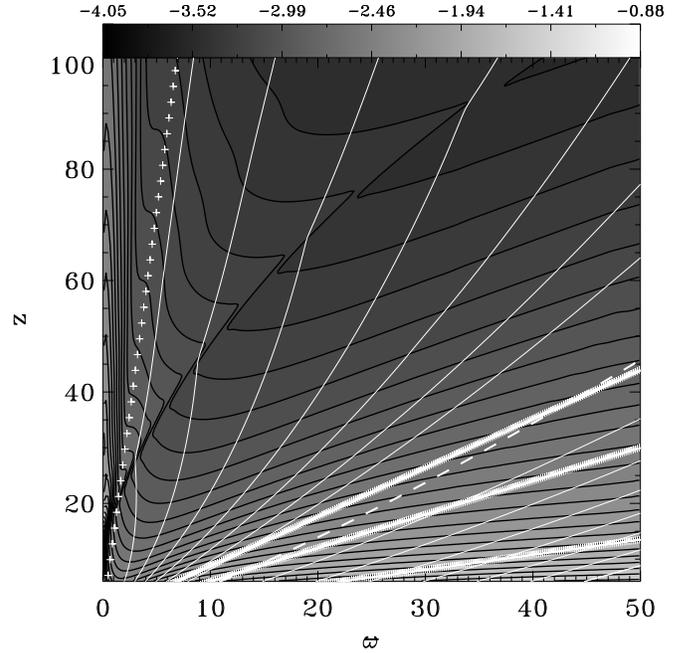}}
\caption{Logarithmic density contours (black lines) and poloidal
         magnetic fieldlines (white thin lines) are plotted for the
         final steady-state reached for model $3DB$. The critical
         surfaces of the initial configuration are given by the
         plus signs; going counterclockwise we find the slow
         magnetosonic and slow magnetosonic separatrix (which
         practically coincide), the Alfv\'en, the fast magnetosonic,
         and the fast magnetosonic separatrix surface (FMSS, close
         to the axis). For the final state, the same surfaces (apart
         from the FMSS) are indicated with the thick white lines.
         However, since the analytical equilibrium is preserved in
         that region, they can hardly be distinguished from those
         predicted by the steady-state analysis. On the other hand,
         the FMSS of the final state is not conical any more but
         diverges from the axis (its position is at the break of the
         magnetic fieldlines and at the discontinuity of the
         density). The dashed poloidal fieldline has been selected
         to compute the integrals of motion shown in
         Fig.~\ref{fig:integrals}.}
\label{fig:bound}
\end{figure}

\begin{figure}
\resizebox{\hsize}{!}{
 \includegraphics{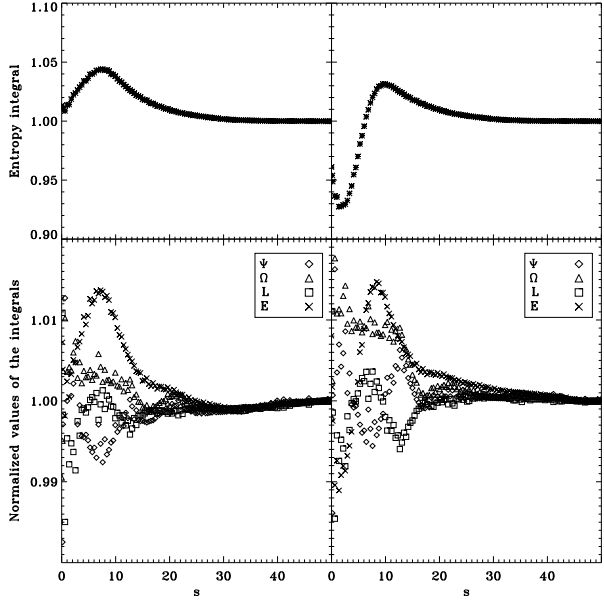}}
\caption{The integrals of motion (Eqs.
         [\ref{eq:intp}]-[\ref{eq:inte}]), normalized to unity,
         are plotted along the poloidal magnetic fieldline (dashed
         line) indicated in Fig.~\ref{fig:bound} for the models
         $1DR$ (left) and $2DB$ (right).}
\label{fig:integrals}
\end{figure}

In all cases the initial configurations are maintained and are
almost identical: density contours and magnetic fieldlines are
being plotted for model $3DB$ (same as model $2DB$ but in higher
resolution) in Fig.~\ref{fig:bound}. The surfaces corresponding
to the slow magnetosonic and slow magnetosonic separatrix surface,
the Alfv\'en surface, and the fast magnetosonic surface of the
final steady state solution are perfectly coinciding with those
of the initial analytical solution. This correspondence was also
found in GVT06, but it was not so tight: this may be most likely
ascribed to the different numerical codes used for the
simulations. Conversely, the numerically reshaped fast
magnetosonic separatrix (FMSS) diverges from the analytical
conical position, and, as will be discussed later, corresponds
to a weak shock which can be seen at the density jump and in the
break of the poloidal fieldlines.

In Fig.~\ref{fig:integrals}, the integrals of motion (Eqs.
[\ref{eq:intp}]-[\ref{eq:inte}]) are being shown for models $1DR$
and $2DB$ respectively, normalized to their outer boundary values.
The fieldline along which they are computed, begins in a region
where $V_z < V_{Alfv\acute{e}n;z}$ while it crosses the Alfv\'enic
and fast classical critical surfaces. This is particularly important
since a fieldline with its rooting point already in the
super-Alfv\'enic or the super-fast region would give much more
similar and hence misleading results for the two cases. It is clear
that the integrals of the final solution deviate by less than
$\sim 1\%$ when compared to the theoretically expected one, with
the exception of the entropy integral $Q$, which shows a rather more
sensitive behavior. Note an improvement on the constancy of those
integrals as compared to GVT06, where they were found within
$\lesssim 15\%$ of their analytical values, again with the exception
of $Q$.

Therefore, Fig.~\ref{fig:integrals} -- along with the fact that
both models $1DR$ and $2DB$ give identical final outcomes -- suggests
that the results obtained are not really sensitive to the number of
boundary conditions imposed. The reason for this, is the fact that
the solution is topologically stable and remains close to the initial
one. Nevertheless, in order to be physically consistent, the correct
number of boundary conditions has to be specified always.

\subsubsection{Shock at the FMSS}

We now adopt the setup of $2DB$ and perform simulations by
effectively increasing the resolution to examine both the behavior
and nature of the density and pressure jump observed (models $4DH$,
$5DH$, $6DH$).

\begin{figure}
\resizebox{\hsize}{!}{
\begin{tabular}{cc}
 \includegraphics{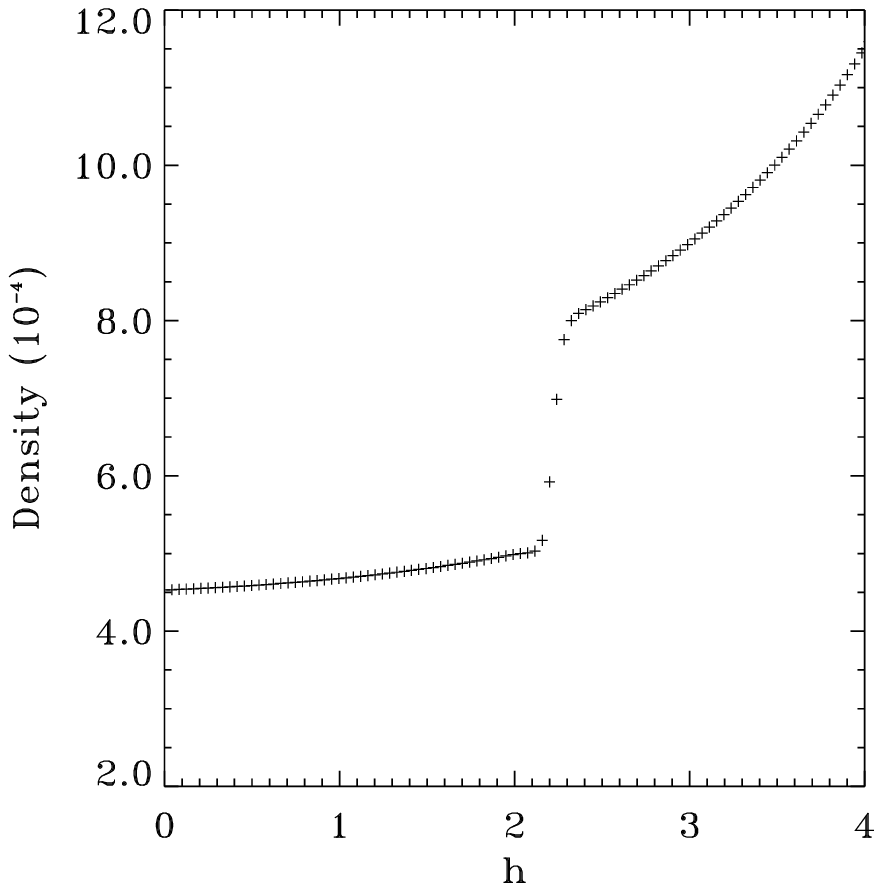} &
 \includegraphics{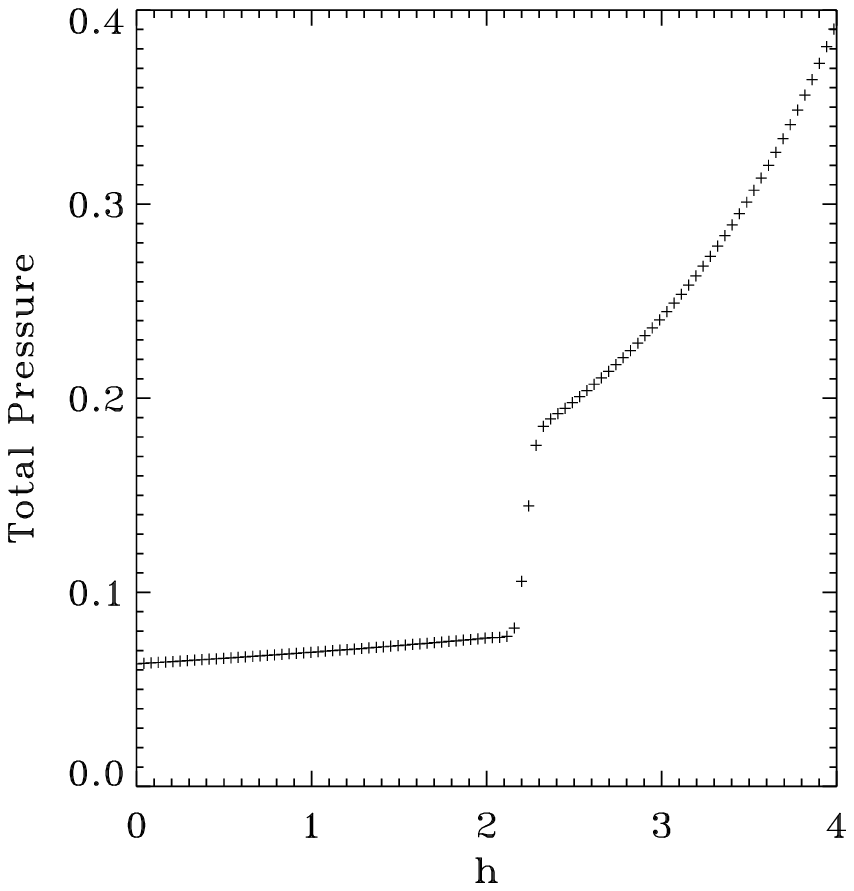}  \\
\end{tabular}}
\caption{The density (left) and total pressure (thermal plus magnetic,
         right) jumps are being plotted for model $6DH$ along the
         perpendicular to the shock direction centered at the
         point $x = 5, y = 33$ (indicated with the diamond in
         Fig.~\ref{fig:charac}).}
\label{fig:shock}
\end{figure}

\begin{figure}
\resizebox{\hsize}{!}{
 \includegraphics{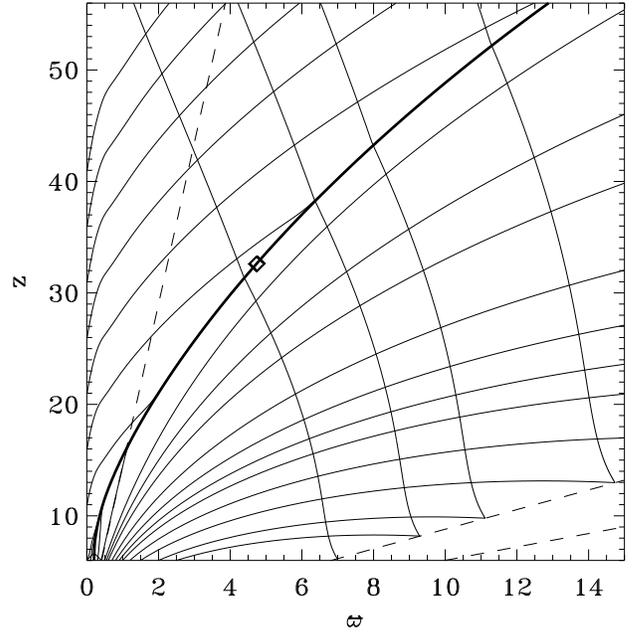}}
\caption{The characteristics (thin solid lines) of the fast magnetosonic
         waves are plotted in a zoomed region around the shock (thick
         line) for model $5DH$. The point indicated with a diamond is
         where the quantities of Fig.~\ref{fig:shock} are plotted. The
         dashed lines are (clockwise) the analytical FMSS, the fast
         poloidal critical surface and the Alfv\'enic one respectively.}
\label{fig:charac}
\end{figure}

\begin{figure}
\resizebox{\hsize}{!}{
 \includegraphics{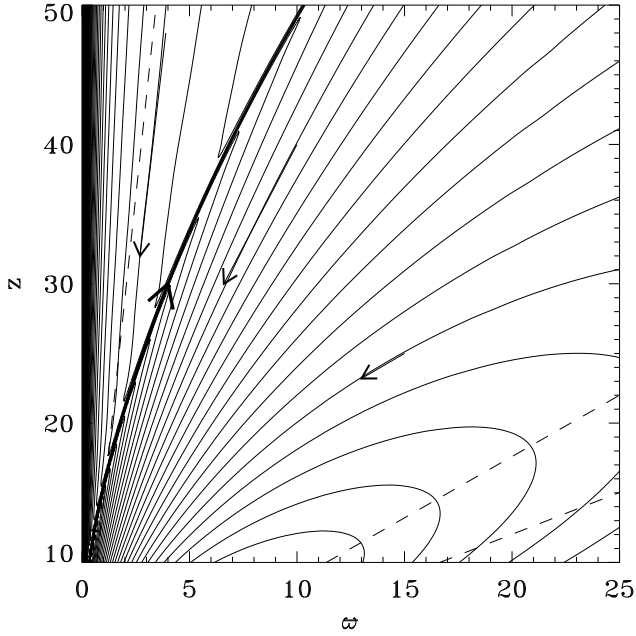}}
\caption{The poloidal currents (loci where $\varpi B_\phi =
         \textrm{const}$) of model $5DH$ are plotted. The arrows
         indicate the direction of the poloidal current density
         $\vec J_p$. A strong current sheet appears tangent to the
         shock. The dashed lines use the notation of
         Fig.~\ref{fig:charac}}
\label{fig:currents}
\end{figure}

The discontinuity manifesting in the simulations of the ADO models is
identified as a weak shock. This conclusion is supported by taking
into account the negative divergence of the velocity (thus denoting
the compressional nature of the discontinuity), the jumps appearing
on the density and pressure and the fact that the gradients steepen
with the increase of the resolution. The density and the total
pressure, thermal plus magnetic ($P + B^2/2$), are plotted across the
shock in Fig.~\ref{fig:shock}. Moreover, the strength of the shock is
found to decrease as we move far away from the base, whereas it
becomes more and more oblique (see Fig. \ref{fig:bound}). Consistent
with the shock is the break of the poloidal fieldlines, as an effect
of the amplification of the tangential component of the poloidal
magnetic field. We remark also that the jump in the entropy is very
small, though, this is expected by the almost isothermal conditions
assumed for the wind ($\gamma =1.05$). However, it has been validated
(see discussion on solutions with different polytropic indexes) that
by increasing $\gamma \rightarrow  \Gamma$ the entropy jump is
increasing as well, as expected in adiabatic conditions.

In Fig.~\ref{fig:charac}, selected fast magnetosonic characteristics
are plotted. Note that the cones on the left of the shock do not
ever cross it but become at best parallel to it. This proves that it
corresponds to the numerically readjusted FMSS of the initial exact
solution. In particular, the model retains its property of a
super-modified-fast solution, and thus a shock develops to preserve
the causal disconnection between the flows downstream and upstream.
Because the cones constructed by the characteristics of the fast
waves in the final superfast region never cross the shock front, it
is not surprising that the sub-modified-fast region is not affected
at all by the modifications taking place at small polar angles. In
other words, the new FMSS behaves like a ``wall'' preventing the
readjustments occurring by the extrapolation close to the axis and
downstream of the analytical FMSS to affect the solution upstream of
the shock. Such a result is supported by all simulations carried out
with the ADO models presented in Table~\ref{table:1}.

Furthermore, in Fig.~\ref{fig:currents} we plot the poloidal
currents for model $5DH$. They are counterclockwise upstream
and far from the FMSS, they change to clockwise very close and
upstream of the FMSS, and finally they change back to
counterclockwise downstream of the FMSS. At the FMSS, where the
azimuthal magnetic field is discontinuous, we have a strong
current sheet with the surface current heading upwards being
tangent to the shock. Thus, the currents seem to have a
``lightning'', or reverse ``N'' shape, with their middle part
being parallel to the shock. This distribution of the currents,
and in particular the direction of the resulting
$\vec J_p \times \vec B_\phi$ force, is consistent with the
decollimation and deceleration that the flow experiences as it
passes through the shock. Thus, one of the effects of the new
FMSS is to bend the streamlines away from the z-axis avoiding
the over-collimation property of the original analytical
solution. The collimation and decollimation processes that can
be derived from such a plot are also discussed in GVT06 (see
Fig.~6 there).

\subsubsection{Smoothening the flow near the rotational axis}

\begin{figure}
\resizebox{\hsize}{!}{
 \includegraphics{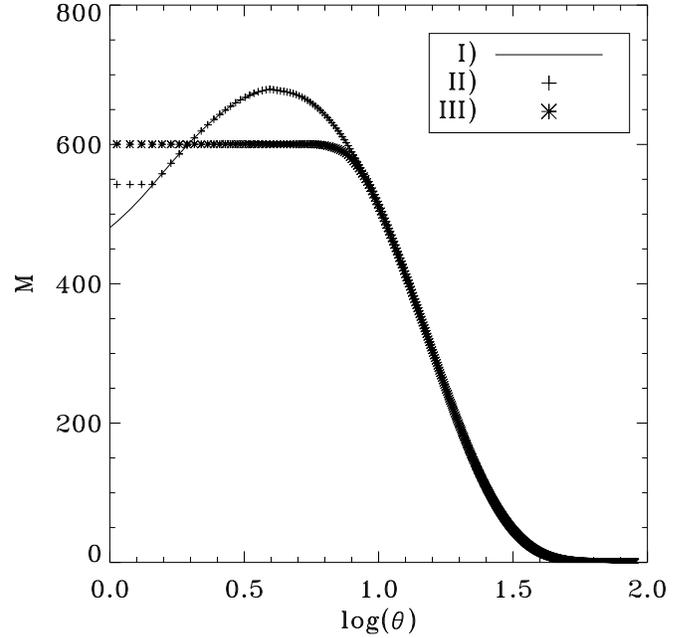}}
\caption{Plot of the initial alfv\'enic mach number, $M$, vs log($\theta$)
         with the three extrapolation schemes assumed for models I) $2DB$,
         II) $7DS$ and III) $8DS$.}
\label{fig:m}
\end{figure}

We now proceed to study how different types of extending the
solution up to the axis affect the final outcome of the simulation.
We adopt the following simple, but diverse enough, extrapolation
schemes for the key functions which are shown in Fig.~\ref{fig:m}
as well:
\begin{equation}
f(\theta) = \left\{
\begin{array}{rll}
 \textrm{  I) ($2DB$)} & \textrm{linear extrap.}
               & \textrm{for } \theta < \theta_{min} \\
 \textrm{ II) ($7DS$)} & \textrm{flat extrap.}
               & \textrm{for } \theta < \theta_{min} \\
 \textrm{III)  ($8DS$)} & \textrm{smooth extrap.}
               & \textrm{for } \theta_{FMSS} < \theta \lesssim 7.5^{\circ}
\end{array}
\right.
\end{equation}
Notice that case I) is applied to all but $7DS$ and $8DS$
numerical models of the ADO solution presented in this paper.

\begin{figure}
\resizebox{\hsize}{!}{
 \includegraphics{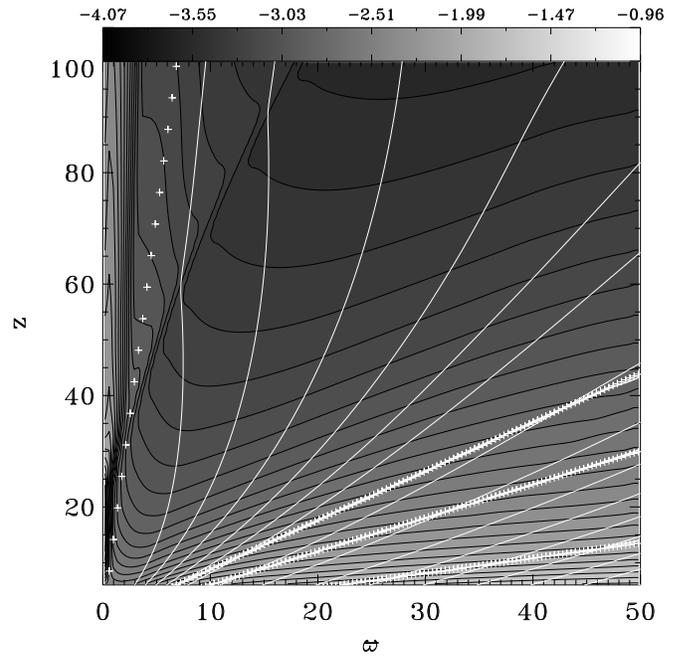}}
\caption{Logarithmic density contours and magnetic poloidal fieldlines
         for model $8DS$  (case III). The symbols are the same as in
         Fig. \ref{fig:bound}. The results for model $7DS$ (case II) are
         identical with those of models $2DB$ the morphology of which
         can be seen in Fig.~\ref{fig:bound}.}
\label{fig:sing}
\end{figure}

\begin{figure}
\resizebox{\hsize}{!}{
 \includegraphics{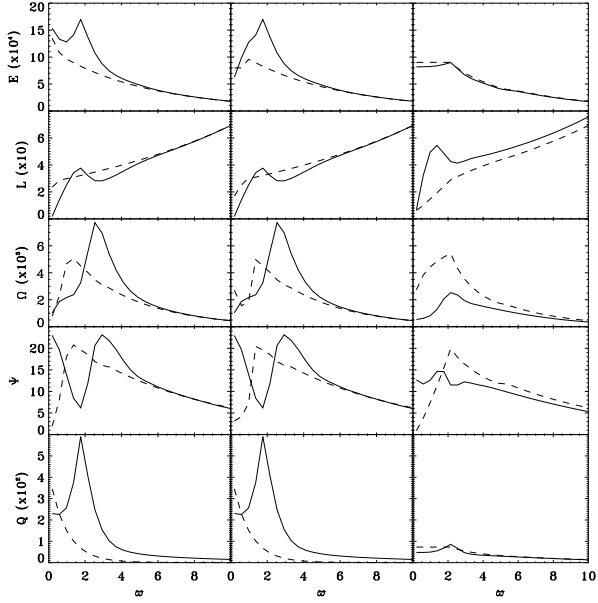}}
\caption{The integrals of motion (Eqs. [\ref{eq:intp}]-[\ref{eq:inte}])
         are plotted along $\varpi$ at $z = 50$ for cases I) 2DB (left),
         II) 7DS (middle) and III) 8DS (right). The dashed line
         represents the initial setup whereas the solid the final
         steady-state reached.}
\label{fig:invar}
\end{figure}

Fig.~\ref{fig:sing} displays the contours of the density for case III)
after the simulation has reached a steady state, whereas cases I) and
II) give identical results which can be seen in Fig.~\ref{fig:bound}.
These first two schemes, which involve modifications only for a small
polar angle, display very similar features, despite their different
extrapolation assumption. Case III) is of particular importance since
the initial solution is sub-modified-fast, i.e. the whole domain is
causally connected. However, the shock is still present in the final
steady-state reached. Such a result suggests that even by starting from
an analytical solution that does not cross the modified-fast critical
point, it will probably self-adapt to an ``astrophysically correct''
solution. With such a term, we mean that it will be consistent with the
causal disconnection -- of the launching and termination regions --
expected to exist in the jet phenomenon. This was somehow anticipated
because, for the super-modified-fast extrapolation schemes of cases I)
and II), the numerical FMSS is encountered before the analytical one
along the flow.

On the other hand, these simulations demonstrate the stability and
robustness of the analytical solution since the classical poloidal
critical surfaces are not readjusted at all for any of these cases. We
conclude that the final steady-state numerical solutions do not show
any sensitivity neither on the extrapolation scheme nor to an initial
sub-modified-fast solution as far as the criticality condition and the
upstream of the shock domain are concerned.

Finally, in order to argue that the extrapolation schemes adopted
do not correspond to any physical inconsistencies, we present in
Fig.~\ref{fig:invar} the invariants (Eqs.
[\ref{eq:intp}]-[\ref{eq:inte}]) along $\varpi$ at $z = 50$ for each
case for the initial and final states. Note that case I), which is
applied in almost all simulations, is particularly interesting since
the inner region is naturally substituted by a lower $\Omega$
mimicking an outflow coming from a slower rotating star.

\subsubsection{Lower boundary}

We now consider the influence of the choice of $z_{min}$ to the
final steady-state reached since the origin is a point where several
variables become singular. We construct models $9Dz$ and $10Dz$, by
lowering and increasing $z_{min}$ respectively, and we carry out
simulations in order to investigate this issue.

\begin{figure*}
\resizebox{\hsize}{!}{
\begin{tabular}{cc}
 \includegraphics{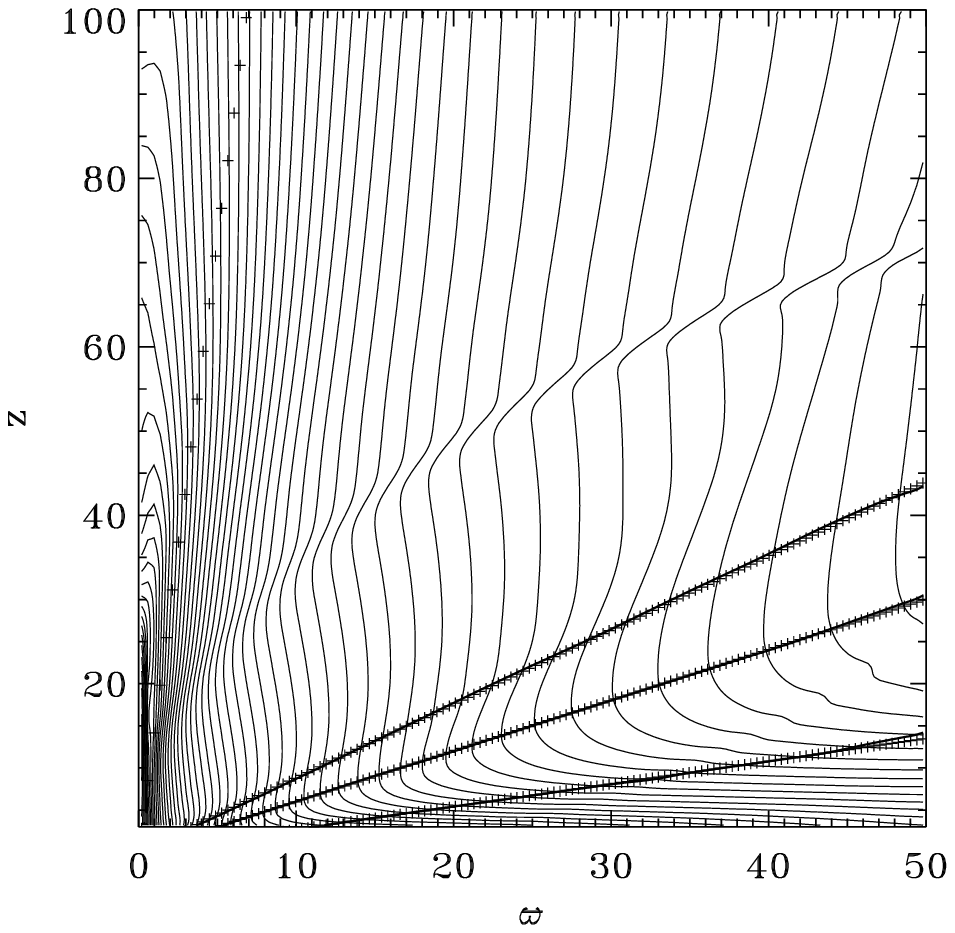} &
 \includegraphics{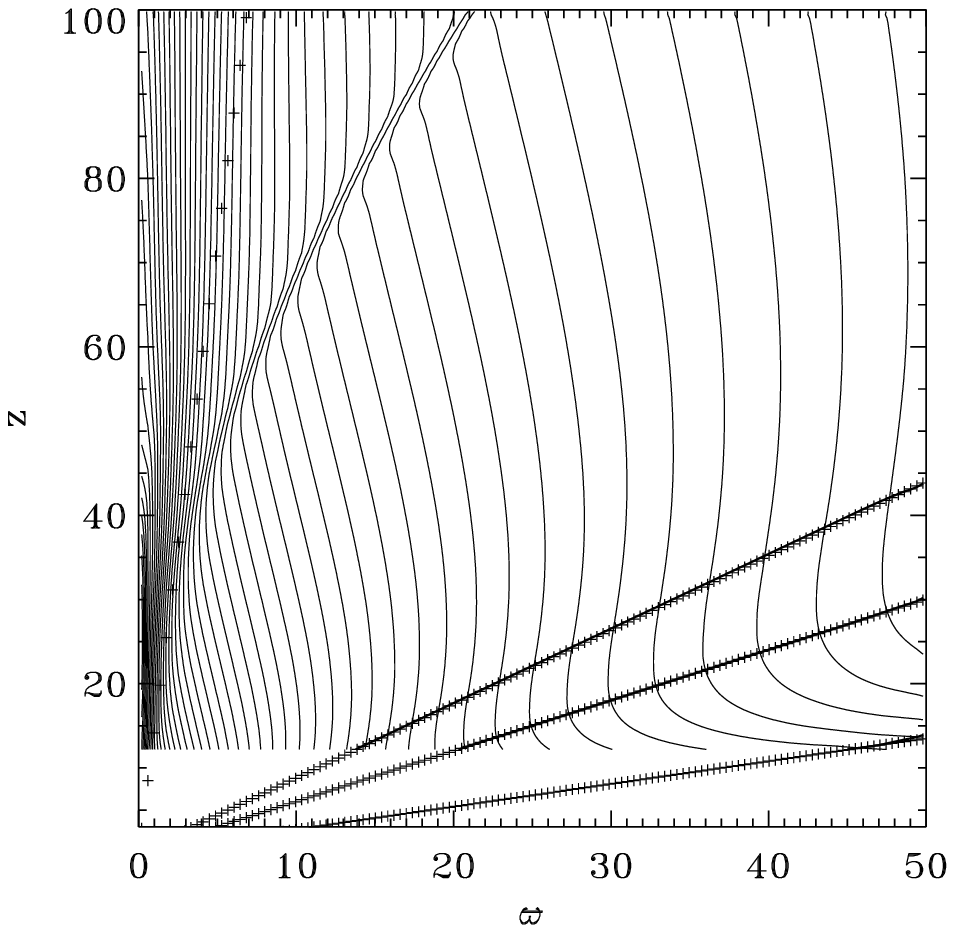}\\
\end{tabular}}
\caption{Contours of the logarithm of the pressure are being plotted
         for models $9Dz$, i.e. $z=3$ (left) and $10Dz$, i.e. $z=12$
         (right). The values increases towards the axis from a minimum
         value of -4.0 to a maximum of 0.0, while the iso-baric lines
         differ by 0.1. The initial and final poloidal critical surfaces
         are also indicated with the notation already mentioned in
         Fig.~\ref{fig:bound}.}
\label{fig:zmin}
\end{figure*}

Fig.~\ref{fig:zmin} shows the pressure contours of models $9Dz$
(left) and $10Dz$ (right); the position of the shock is also evident.
Note that in an iso-density contour plot, the shock would not be
clearly distinguished because the discontinuity appearing in model
$9Dz$ is approximately aligned to the iso-density surfaces, as seen
in Fig.~\ref{fig:bound}. Even though the classical critical surfaces
of both cases do not present any deviation from the initial model,
the region where the shock front develops is considerably different.
This result can be understood by noting the following. The
characteristics of the fast magnetosonic waves are directly related
to the formation of the shock. Therefore, the values of the physical
variables in the extrapolated region control its position. Recall
that these values are kept fixed since we are in the super-modified
fast region. So, we speculate that the closer to the origin we set
the bottom boundary, the stronger these extrapolated values will be
affected by the steep gradients of the singular origin and the more
the characteristics defining the shock will deviate from the
analytical FMSS.

\subsubsection{Extension to the equator and energetics}

We proceed now to extend GVT06 down to the equator and simultaneously
examine the effect of different kinds of energy input/output. To
achieve the former we adopt spherical coordinates in order to
naturally avoid the singularity at the origin, where all critical
surfaces coincide. The simulations are performed in the grid with
$R\in[10, 90]$ and $\theta \in [0, \pi/2 - \epsilon]$, utilizing
$R\times\theta = [408\times 256]$ uniform zones, where
$\epsilon \simeq 1^\circ$. We effectively change the way the energy
equation is treated by implementing and evolving the following 3
setups: i) applying $\gamma = 1.05$ ($11DE$), i.e. according to the
analytical solution, ii) assuming $\Gamma = 5/3$ ($12DG$) to
examine the solution's behavior to an adiabatic evolution and
iii) constraining the system under an isothermal condition ($13DI$).

\begin{figure*}
\resizebox{\hsize}{!}{
\begin{tabular}{ccc}
 \includegraphics{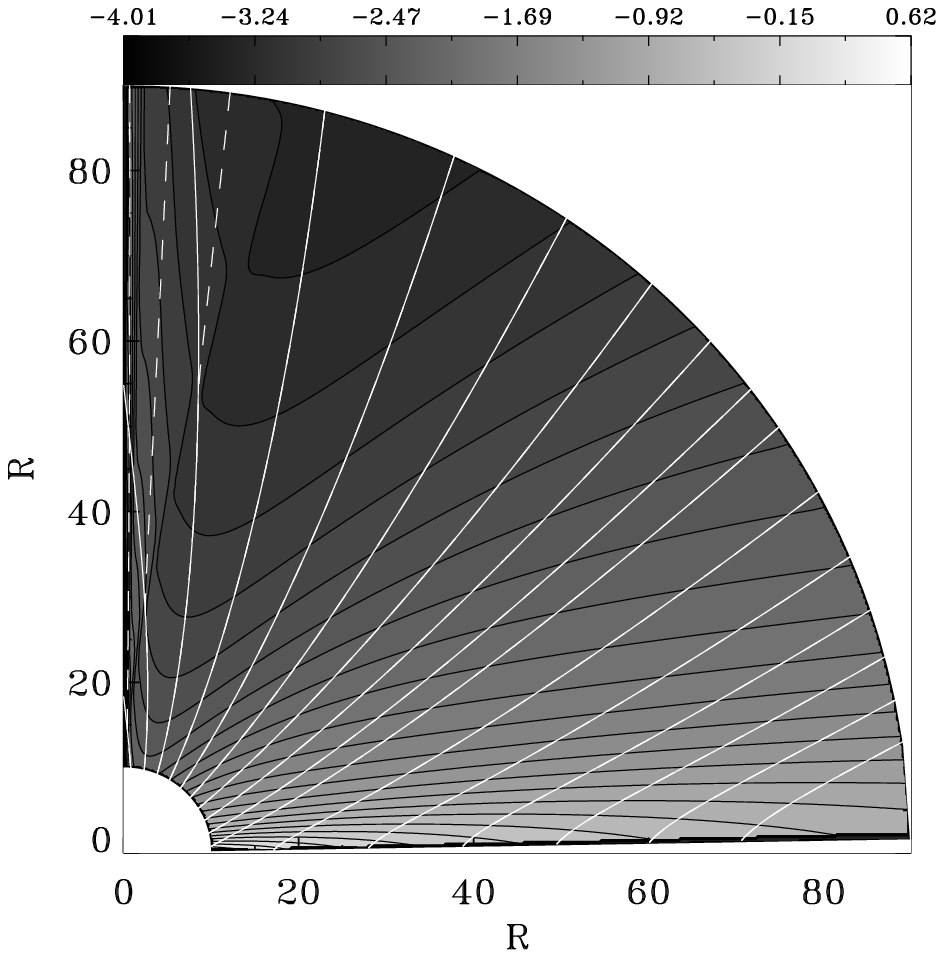} &
 \includegraphics{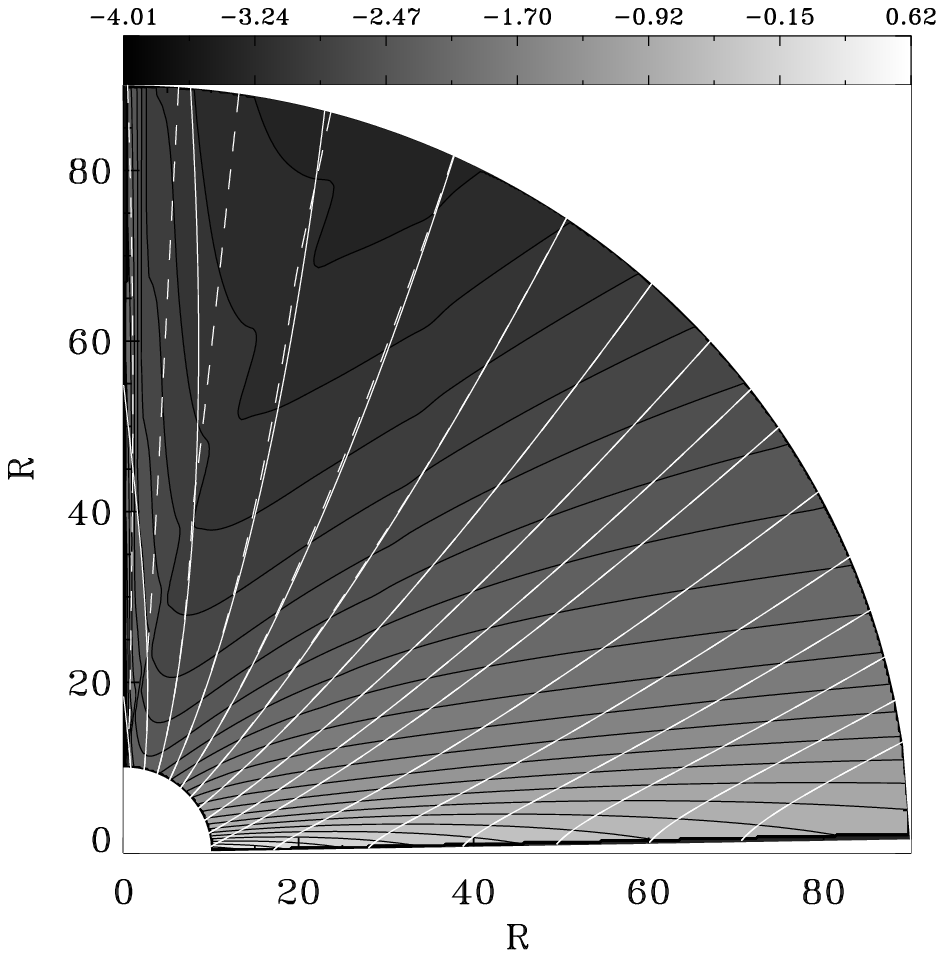} &
 \includegraphics{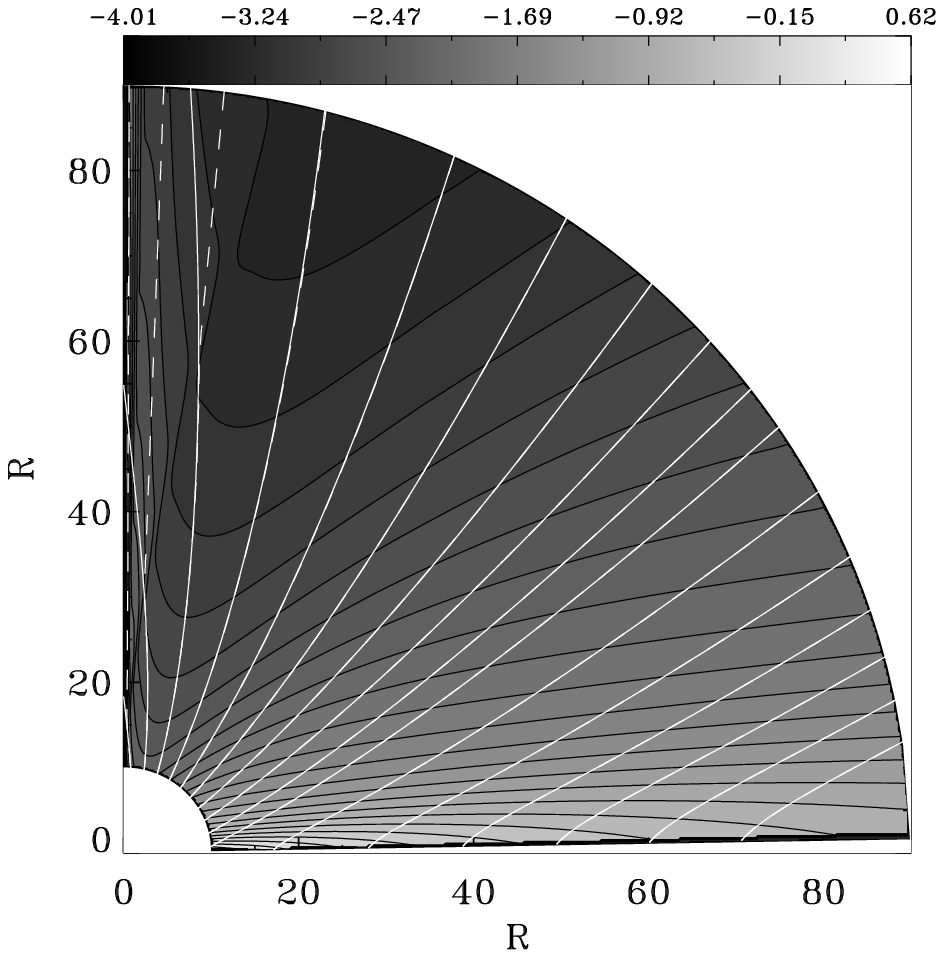}\\
\end{tabular}}
\caption{Logarithmic contours of the density (black lines) are being
         plotted for models $11DE$ (polytropic evolution, left), $12DG$
         (adiabatic evolution, middle) and $13DI$ (isothermal evolution,
         right). The initial (solid) and final (dashed) fieldlines are
         sketched as well (white lines).}
\label{fig:spher}
\end{figure*}

The left panel of Fig.~\ref{fig:spher} gives the iso-density
surfaces and the shape of selected magnetic fieldlines of model
$11DE$. As expected, including the equator does not involve any
new phenomena, since the analytical solution is well defined
there. However, we note that such a computational domain, along
with the fixed boundary conditions imposed there, is a physically
more consistent choice to describe the MHD outflow, justified by
the geometric properties of the disk-star system. Finally, notice
that the position of the shock is consistent with the analysis we
performed with the cylindrical coordinates, with the role played
by $z_{min}$, being now by $R_{min}$.

On the other hand, interesting results are displayed on the middle
and right panels of Fig.~\ref{fig:spher}, where selected fieldlines
are plotted for the initial setup (solid) and the final numerical
steady-state solution (dashed). The plots indicate that even though
the polytropic relation is an unavoidable simplification to derive
the exact solution of VTST00, it is found to contribute negligibly
on the final outcome reached by the numerical evolution. Besides,
this is not surprising since the necessary distribution of the
pressure needed for constructing such solution is still the same
for these cases, with only the energetics during time evolution
being different. Such a pressure is crucial in order to force the
Alfv\'enic critical surface to be close to the equator. For a
thorough discussion see Ferreira \& Casse (\cite{Fer04}).

\subsubsection{Long term evolution}

\begin{figure}
\resizebox{\hsize}{!}{
 \includegraphics{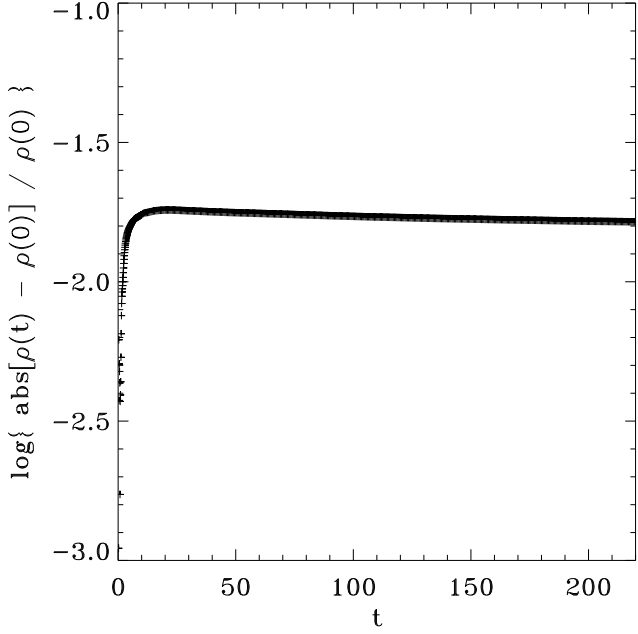}}
\caption{Normalized density deviation from its initial value
         as a function of time at the point $(\varpi,z) = (10,35)$.}
\label{fig:time}
\end{figure}

Finally, the structural stability argument has to be made concrete
by evolving the solution for larger time scales. For that reason we
constructed model $14DT$, a case identical to $2DB$, but with its
right edge further extended to avoid spurious reflections at the
boundaries. Moreover, we choose a point in the upstream of the shock
region, i.e. in the sub-modified-fast domain, and plot the
deviations of the density from its initial value as a function of
time.

It is evident from Fig.~\ref{fig:time} that initially ($t \lesssim 2$)
the sub-modified-fast solution is being perturbed due to the proper
modifications at the axis. However, it converges quickly ($t < 5$) to
a constant value, roughly $\sim1.5\%$ different from its initial one.
For the rest of the simulation ($5 < t < 220$) the steady state is
perfectly preserved proving its stability. Though, for $t > 220$,
boundary effects of the imposed outflow conditions of the rightmost
and upper boundary have propagated throughout the domain and start
to artificially affect the solution. It is worth noting here that
the Alfv\'en velocity is of the order of unity at the lower central
region of the computational box. Hence, the Alfv\'en waves have time
to propagate many times before the simulation is brought to a halt.

\subsection{The Analytical Stellar Outflow (ASO) solution}

The meridionally self-similar solution is in general less
complicated to study, since it does not involve any separatrix
in the super-Alfv\'enic region. Therefore, we will mainly
investigate here its topological stability and its response
towards different restrictions on the evolution of its energy
equation. Notice that models $1SR$, $2SL$ and $3SL$ evolve in
time with the analytically derived source term participating
throughout the whole computational domain. For the rest of
the ASO models investigated, the details are given in the
following.

\begin{figure}
\resizebox{\hsize}{!}{
 \includegraphics{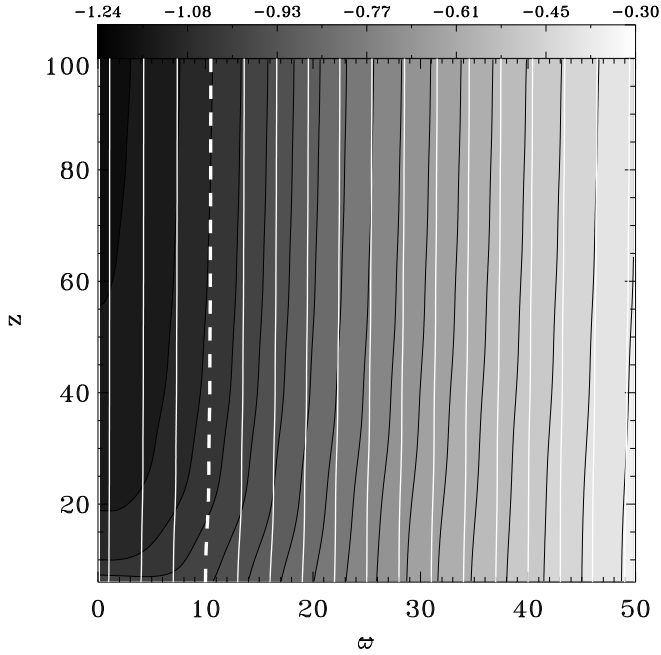}}
\caption{Logarithmic density contours (black lines) are sketched
         for the final time of the simulation for the model $1SR$ which
         does not include a sub-Alfv\'enic region. The magnetic
         fieldlines are drawn with white lines. The thick dashed one, is
         chosen to plot along the integrals of motion
         (Eqs. [\ref{eq:intp}]-[\ref{eq:inte}]; Fig.~\ref{fig:tssint}).}
\label{fig:tss}
\end{figure}

\subsubsection{Asymptotic configuration}

In Fig.~\ref{fig:tss} we plot the iso-density surfaces along
with selected magnetic fieldlines for a super-Alfv\'enic wind
(model $1SR$). Note that in this case we neither have the
entropy integral nor the energy one, since for the former there
is not any polytropic relation assumed, while for the latter,
an explicit energy source term is participating. The initial
setup is an exact solution throughout the whole computational
domain and hence the final time of the simulation is
arbitrarily chosen to be equivalent to the one selected for
model $1SR$. The stability of this class of solution can be
easily seen in Fig.~\ref{fig:tssint} by the fact that the
integrals of motion deviate by only $\lesssim 0.2\%$.

\begin{figure}
\resizebox{\hsize}{!}{
 \includegraphics{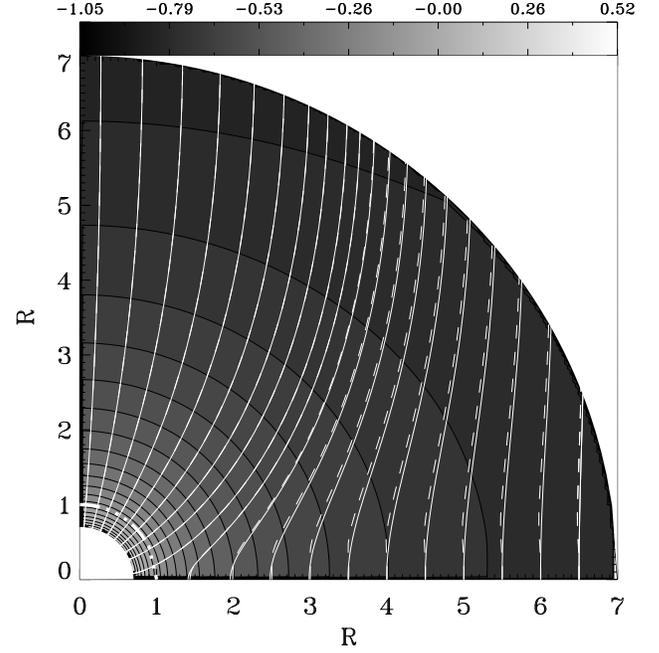}}
\caption{Selected magnetic fieldlines (white lines) are
         plotted for model $3SL$ which includes a sub-Alfv\'enic
         region. Initial analytical solution (solid) and final
         outcome of the simulations (dashed). The Alfv\'en critical
         surface is indicated with a thick dotted-dashed line. The
         gradients of gray represent the logarithm of the density.}
\label{fig:logar}
\end{figure}

The solution stability has also been successfully validated
throughout all its radial range by adopting a logarithmic
grid in the radial direction in spherical coordinates (models
$2SL$ and $3SL$). Note that model $3SL$ is particularly
interesting since $R\in[0.7, 7.0]$ with $R_* = 1$, which
implies that both sub-Alfv\'enic and super-Alfv\'enic regions
are consistently included inside the computational box.
Fig.~\ref{fig:logar} gives the magnetic fieldlines of the
final numerical solution which remains identical to the
initial analytical one. Therefore, the application of a
logarithmic grid allows the approach of the stellar surface,
and hence the fixed boundaries imposed are physically
justified.

\subsubsection{Energetics}

It has already been mentioned that one of the main questions
posed by the mixing of the two self-similar solutions is on the
treatment of the energy equation. Although this will be discussed
extensively in a companion work, we carry out simulations here
to examine whether the ASO model can reach a steady-state when
different energy input/output is included. We firstly address the
super-Alfv\'enic outflow assuming, as in the previous ADO
solution, $\gamma = 1.05$ (model $4SP$) or the adiabatic case
$\Gamma = 5/3$ (model $5SG$).

\begin{figure*}
\resizebox{\hsize}{!}{
\begin{tabular}{ccc}
 \includegraphics{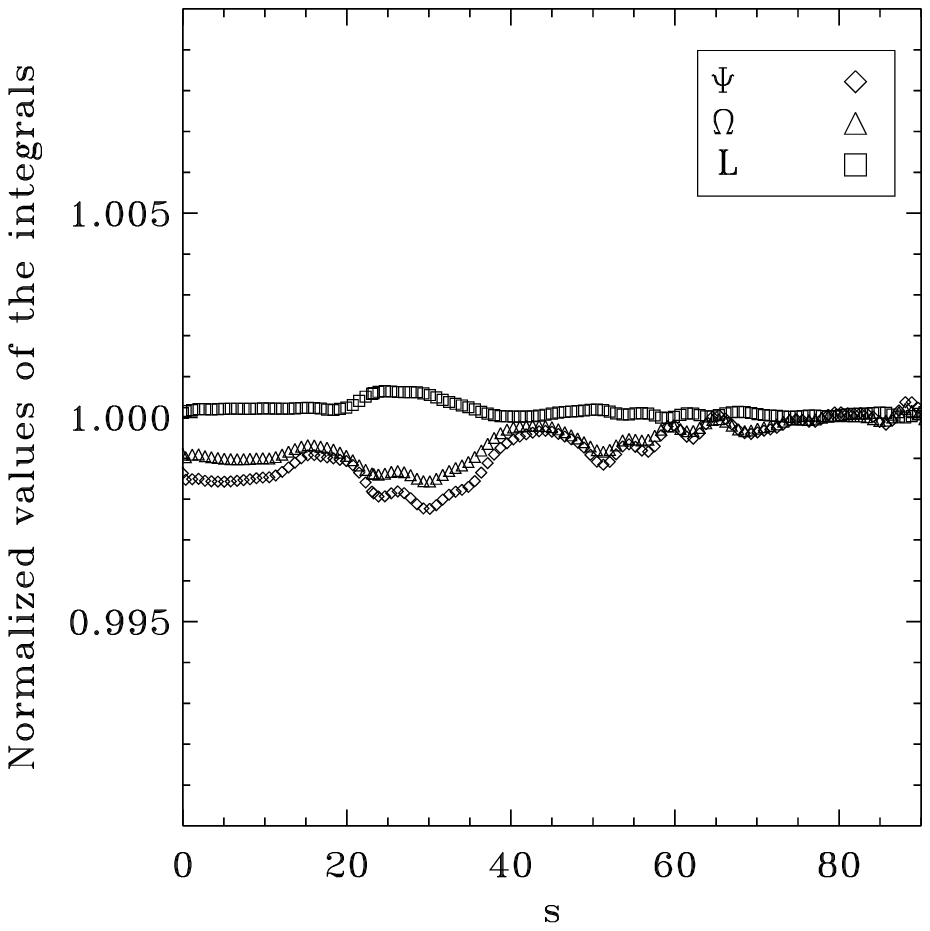} &
 \includegraphics{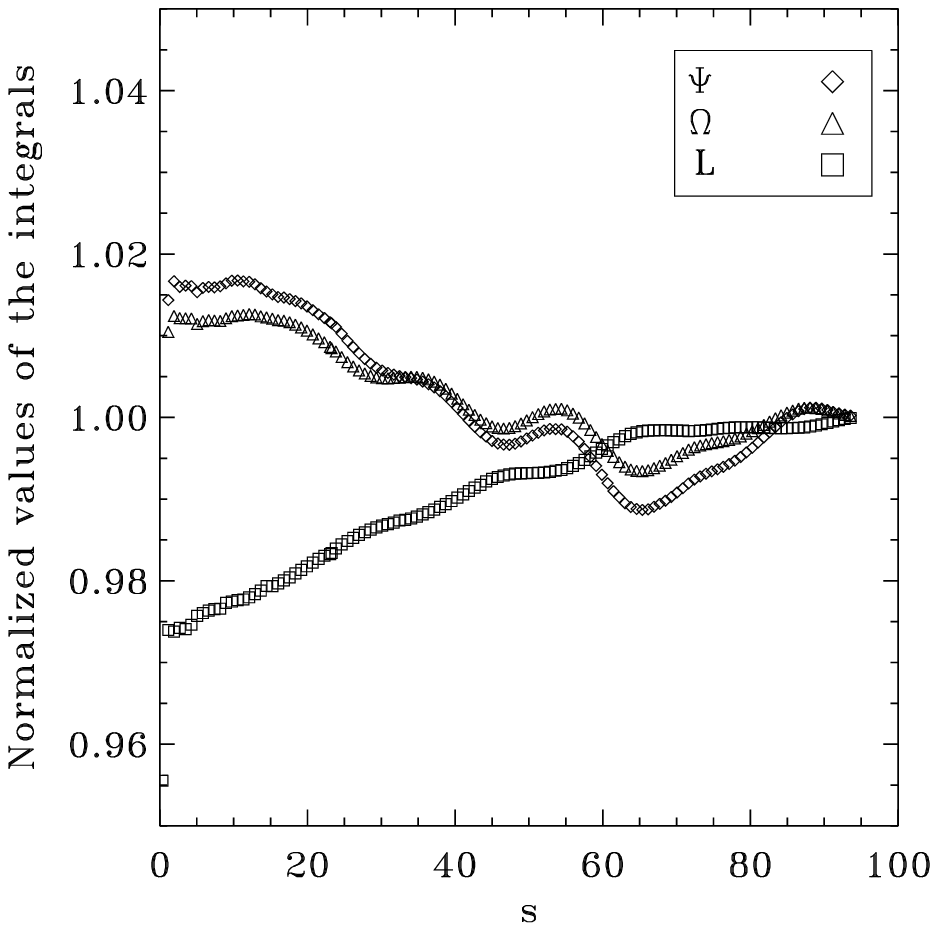} &
 \includegraphics{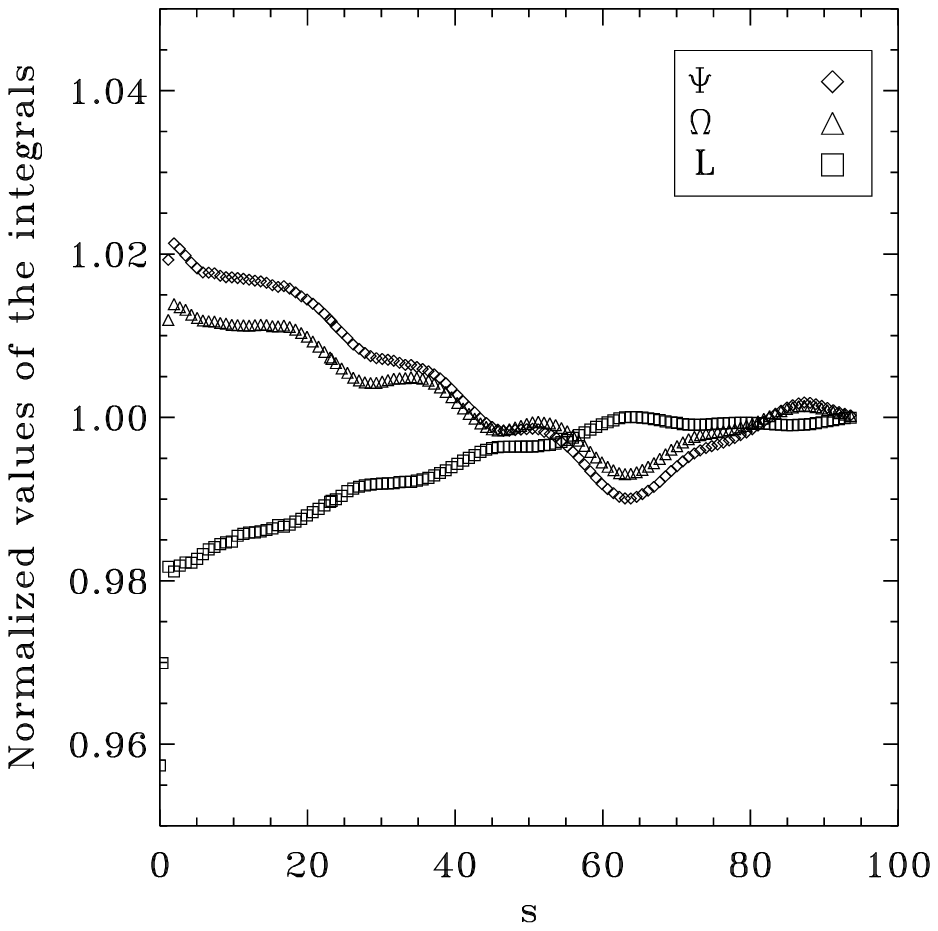} \\
\end{tabular}}
\caption{The integrals of motion are plotted along the indicated
         fieldlines (white dashed line in Fig.~\ref{fig:tss}) for
         models $1SR$ (left), $4SP$ (middle) and $5SG$ (right).}
\label{fig:tssint}
\end{figure*}

\begin{figure}
\resizebox{\hsize}{!}{
 \includegraphics{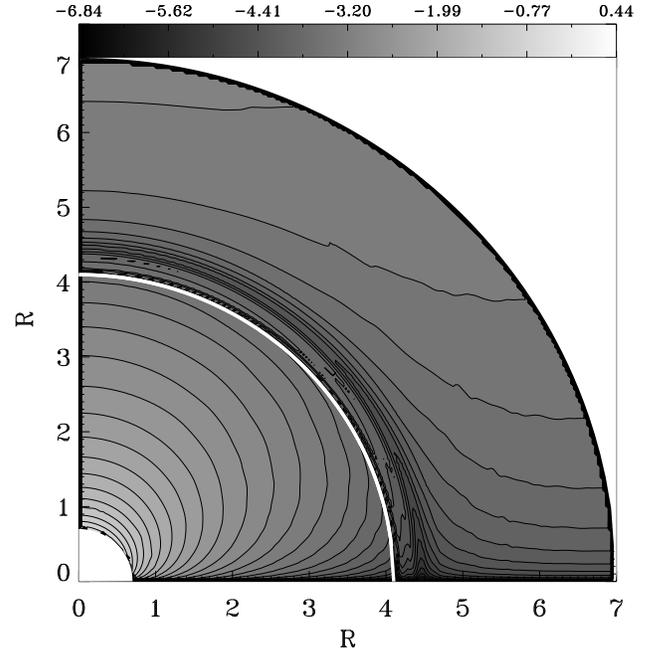}}
\caption{The logarithm of the absolute value of the heating distribution
         is being plotted (model $3SL$). The thick white line indicates
         the surface after which the energy input changes sign and becomes
         cooling. Note that in the inner region, the heating distribution
         gradually decreases by $\sim 4$ orders of magnitude, while
         in the outer one by only half.}
\label{fig:heat}
\end{figure}

The outcome of the simulations is a quasi-steady-state that remains
very close to the initial solution. With the term quasi-steady, we
imply the state when the timescale of the evolution of the system
is much larger as compared to the one of the initial more intense
readjustments. Especially at the first time steps, the solution is
strongly perturbed searching for a new equilibrium due to the
different energy source terms imposed. Later on, after the system
relaxes, the outer radial regions are found to have remained almost
unmodified, whereas, close and along the axis the pressure has
increased by roughly half an order of magnitude followed by a
slight decrease of the density. This is not surprising, since the
analytical explicit energy source term, being taken into account in
model $1SR$, is of negative sign there, i.e. corresponds to
energy loss. On the contrary, the polytropic case corresponds to a
positive input of energy while model $5SG$ to a zero
heating/cooling. Hence, the pressure keeps increasing in both cases,
due to the absence of the needed cooling, until it reaches a new
numerical quasi-equilibrium configuration. The quasi-steady-state
reached can be judged by Fig.~\ref{fig:tssint}, where the integrals
of motion show deviations of $\lesssim 3\%$ after $\sim 100$
Keplerian rotations at the Alfv\'en radius. This is an expected
outcome considering that we are in the super-Alfv\'enic region: we
know that the efficiency of the thermal driving of the flow is
concentrated very close to the base, where almost the whole
acceleration occurs. This can be seen in Fig.~\ref{fig:heat} where
we plot the energy source term coming from the analytic solution
(Eq. [\ref{eq:lam}]). Up to four Alfv\'en radii, the heating
distribution decreases by $\sim 4$ orders of magnitude thus proving
its crucial role accelerating the outflow in the inner region.
Afterwards, when the flow is propagating with its asymptotic speed,
the energetics play a less important role.

One of the properties of these solutions are the oscillations of
the fieldlines. The analytical results  predict this possibility
and we can argue that different energetic processes in the
super-Alfv\'enic region amplifies these features.

\begin{figure*}
\resizebox{\hsize}{!}{
\begin{tabular}{cc}
 \includegraphics{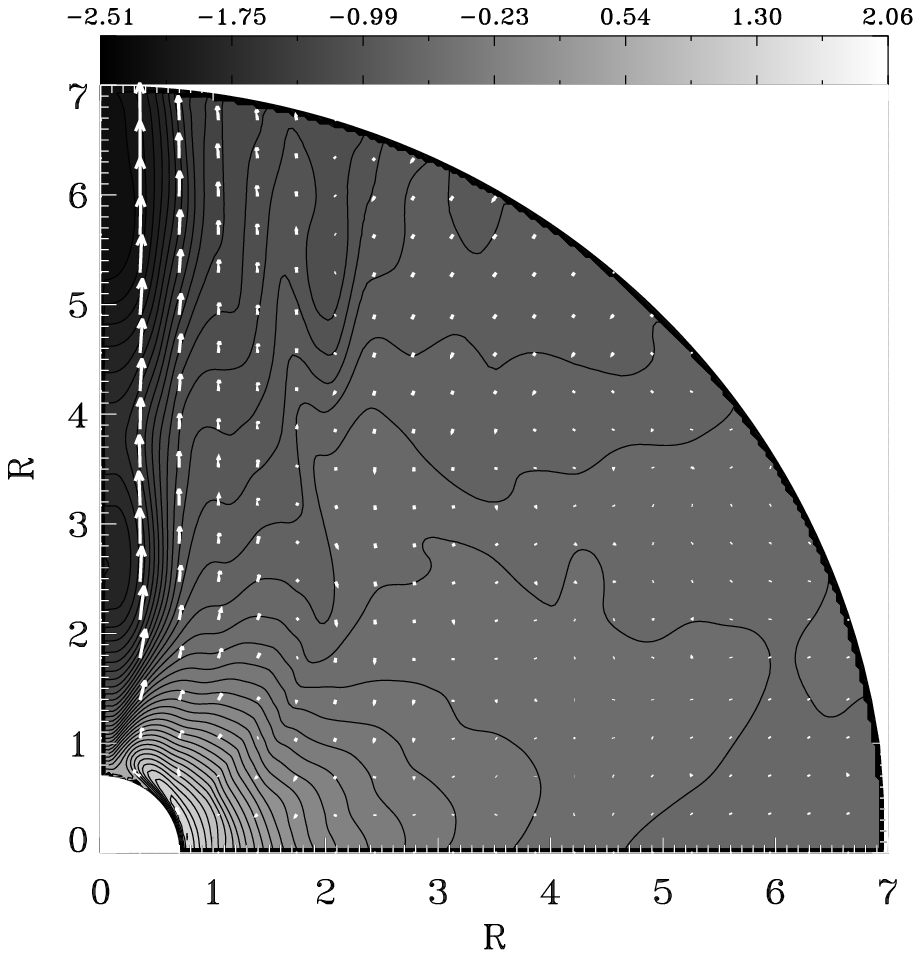} &
 \includegraphics{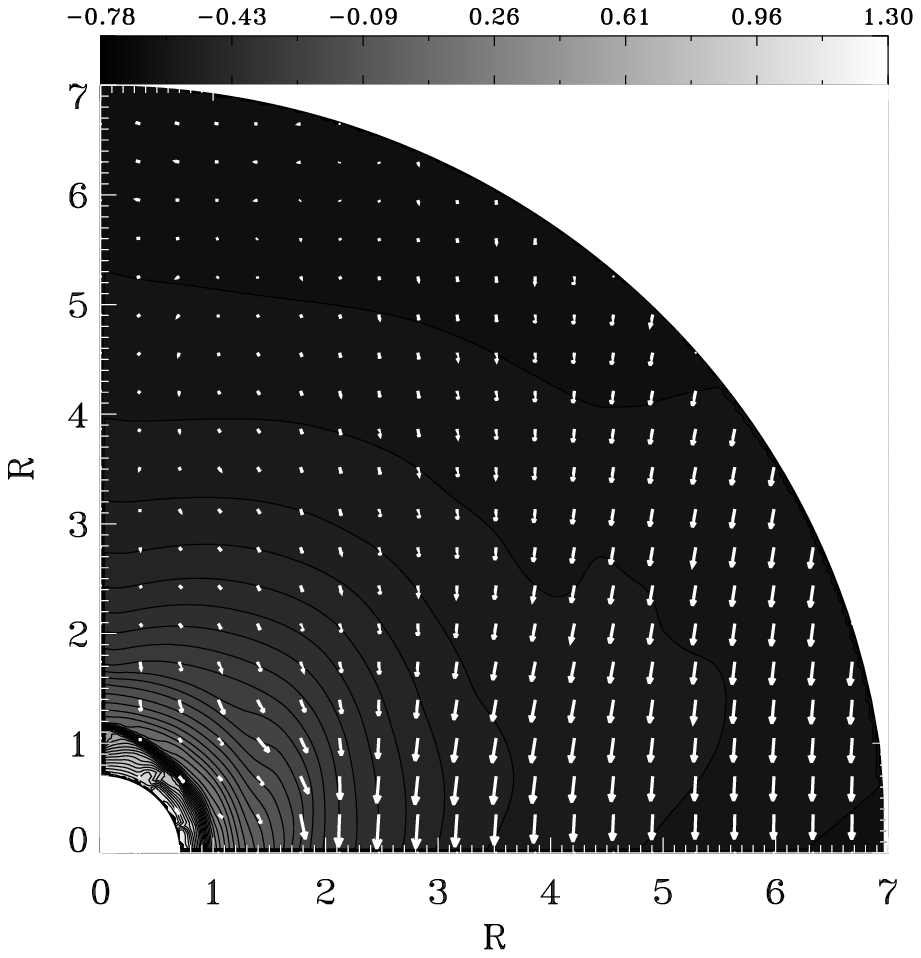}\\
\end{tabular}}
\caption{Logarithmic density contours are being plotted for a polytropic
         evolution ($6SL$ left) and an adiabatic one ($7SL$
         right). The velocity vectors overplotted are in the range
         $[4\cdot10^{-3},4]$ and $[7\cdot10^{-7},5\cdot10^{-1}]$ for the
         polytropic and adiabatic case, respectively. These models never
         reach a steady-state.}
\label{fig:diff}
\end{figure*}

The results are totally different if the same type of simulations,
i.e. polytropic or adiabatic evolution, are performed with the
sub-Alfv\'enic region included (models $6SL$ and $7SL$): in this
case there is no steady-state reached. On the left of
Fig.~\ref{fig:diff} a snapshot of the turbulent evolution is
displayed when the polytropic assumption with $\gamma=1.05$ is
applied. In fact, such a model is able to drive a sporadic low
density outflow around the axis, as it can be seen by the velocity
vectors.

Conversely, an adiabatic evolution, Fig.~\ref{fig:diff} on the right,
forces the system to collapse towards the star, asymptotically
approaching a static atmosphere. This is because the ASO model is
thermally driven and when we impose an adiabatic equation of state
we effectively switch off all the heating needed to drive the
outflow. As expected, the energy processes at the base of
meridionally self-similar winds are crucial for their evolution.

\section{Summary and conclusions}
\label{sec:disc}

In this paper we have studied several physical and numerical aspects
concerning two classes of the self-similar models, each associated
with a disk- and a stellar-wind, in the framework of the upcoming
work that combines them to describe a two-component outflow. These
analytical solutions (ADO and ASO) were appropriately modified,
implemented as initial conditions and evolved in time. Our main
conclusions are the following:

\begin{itemize}
\item
The {\it Analytical Disk Outflow (radially self-similar) solution}
has been successfully validated for its stability and robustness
against several physical and numerical issues. This argument holds
true even though the analytical solution was in many cases
significantly modified. We have constructed numerical models and
carried out simulations a) by assuming the extreme cases of an
isothermal and an adiabatic evolution,  b) by treating the diverging
behavior of the solution at the axis with different kinds of
extrapolation schemes, mimicking a stellar wind component and c)
by changing the size, resolution and geometry of the computational
box.

In all cases, the poloidal critical surfaces, with the exception of
the FMSS, were not readjusted, but rather matched perfectly to their
initial position. The numerical solution always maintained the
property of the successful crossing of all three critical surfaces
producing an outflow causally disconnected from the base. This is
achieved with the formation of a shock, corresponding to the
numerically readjusted FMSS. In particular, this shock acts as a
``wall'' protecting the sub-modified-fast magnetosonic regions
(source regions of the disk wind) from any perturbations taking
place due to the modification of the models close to the axis (e.g.
an effective stellar wind). However, the numerically readjusted
FMSS (shock) does not coincide with the analytical one, with this
departure being dictated by the respective numerical modifications
of the models under consideration. A highly significant result is
the fact that such a conclusion holds true even if we initialize the
simulation with a sub-modified-fast solution, i.e. a solution with
its whole domain causally connected. We found that, during the
simulation, such a numerical model self-adapts to produce a shock
(corresponding to the FMSS), hence no information of the downstream
region can travel back to affect the launching region. This implies
that even MHD outflow solutions, that do not successfully cross all
three critical points, will probably converge to ``astrophysically
correct'' solutions once evolved in time (see also Ferreira
\cite{Fer97}).

On the other hand, the study of GVT06 was successfully extended
down to the equator with the help of simulations using spherical
coordinates. Furthermore, by adopting different assumptions for
the energy source terms, it was shown that the solution is only
slightly and accordingly self-modified maintaining all its well
defined properties. This is in agreement with the fact that the
ADO solution describes essentially a magneto-centrifugally
accelerated outflow.

\item
The {\it Analytical Stellar Outflow (meridionally self-similar)
solution}, which was validated in time-dependent simulations for the
first time, maintained its well-defined equilibrium as expected. Such
conclusion is supported by simulations performed with both the super-
and sub-Alfv\'enic regions included. Quite critical are, contrary
to disk winds, the effects of the energetics in such thermally driven
models. Although different assumptions of the energy equation in the
super-Alfv\'enic domain did not yield any significant modification
of the analytical solution, strong variations of the structure of the
axial outflows are found if modifications of the heating/cooling
mechanisms occur in the initial accelerating region. In particular a
polytropic assumption, mimicking isothermal conditions, would produce
a turbulent weak outflow, while an adiabatic evolution asymptotically
reaches a static atmosphere. We are tempted to relate the heating
intermittency and even a switching off in such an ASO solution with
the observed variability of accretion-driven YSO outflows.

\item
All previous statements hold true while being in perfect agreement
with physically consistent requirements, such as specifying the
correct type of boundary conditions:
a) according to the propagation direction of the MHD waves,
b) the axisymmetry holding around the axis and
c) the constancy of certain physical variables at both a conical
surface close to the equatorial plane and a radial one close to the
origin, implying the presence of an underlying disk and the stellar
surface, respectively. Therefore, the results can safely be trusted
since they are not subject to any artificial forcing.

\item
Last, but certainly not least, is the fact that almost all models
of both classes reached a steady- or quasi-steady-state. In this
context, the upcoming mixing of the two complementary classes of
self-similar solutions in order to study a two-component jet is
well founded and promising. The final numerical model will
incorporate both proposed scenarios of a pressure-driven outflow
(ASO) surrounded by an extended, magneto-centrifugally driven
disk wind (ADO). This task is undertaken in the second paper of
this series.
\end{itemize}

\begin{acknowledgements}

The authors would like to thank the referee Jonathan Ferreira
whose constructive comments and suggestions resulted in a much
better presentation of this work. Fruitful discussions with J.
Gracia and C. Fendt are acknowledged as well. The present work
was supported in part by the European Community's Marie Curie
Actions - Human Resource and Mobility within the JETSET (Jet
Simulations, Experiments and Theory) network under contract
MRTN-CT-2004 005592 and the European Social Fund and National
Resources - (EPEAEK II) - PYTHAGORAS.

\end{acknowledgements}

\end{document}